\documentclass[10pt,letterpaper]{article}
\usepackage[top=0.85in,left=1.6in,footskip=0.75in]{geometry}

\usepackage{amsmath,amssymb}
\usepackage{mathtools}

\usepackage{changepage}

\usepackage[utf8x]{inputenc}

\usepackage{textcomp,marvosym}

\usepackage{cite}

\usepackage{microtype}
\DisableLigatures[f]{encoding = *, family = * }

\usepackage[table]{xcolor}

\usepackage{array}

\usepackage{caption,tabularx,booktabs}
\usepackage{tcolorbox}
\usepackage[hyperref]{}
\newcounter{BoxCounter}
\newtcolorbox[use counter=BoxCounter]{Boxes}[2][] {
fontupper=\footnotesize,
center, 
halign=left,
title=\bfseries\sffamily #2 \hfill Box \theBoxCounter,
#1
}
\usepackage{rotating}
\usepackage{url}
\usepackage{nameref}
\usepackage{fontenc}

\newcolumntype{+}{!{\vrule width 2pt}}

\newlength\savedwidth

\newcommand{\beginsupplement}{%
        \setcounter{table}{0}
        \renewcommand{\thetable}{A\arabic{table}}%
        \setcounter{figure}{0}
        \renewcommand{\thefigure}{A\arabic{figure}}%
     }

\usepackage[aboveskip=1pt,labelfont=bf,font=small,labelsep=period,singlelinecheck=off]{caption}
\makeatletter
\renewcommand{\@biblabel}[1]{\quad#1.}
\makeatother

\usepackage{lastpage,fancyhdr,graphicx}
\usepackage{epstopdf}
\begin{document}
\vspace*{0.2in}

\begin{flushleft}
{\Large
\textbf\newline{\textbf{Feature interpretability in BCIs: exploring the role of network lateralization}}

}
\bigskip

\bigskip
Juliana Gonzalez-Astudillo\textsuperscript{1,*},
Fabrizio De Vico Fallani\textsuperscript{1,*}
\\
\bigskip
\textbf{\textsuperscript{1}} Sorbonne Université, Paris Brain Institute(ICM), Inria Paris, CNRS UMR7225, INSERMU1127, AP-HP Hôpital Pitié-Salpêtrière, Paris,France
\\
\bigskip

%
%
%



* corresponing authors: 
\\julianagonzalezastudillo@gmail.com
\\fabrizio.de-vico-fallani@inria.fr

\end{flushleft}
\bigskip

\section*{Abstract}
\begin{small}
Brain-computer interfaces (BCIs) enable users to interact with the external world using brain activity. Despite their potential in neuroscience and industry, BCI performance remains inconsistent in noninvasive applications, often prioritizing algorithms that achieve high classification accuracies while masking the neural mechanisms driving that performance. In this study, we investigated the interpretability of features derived from brain network lateralization, benchmarking against widely used techniques like power spectrum density (PSD), common spatial pattern (CSP), and Riemannian geometry. We focused on the spatial distribution of the functional connectivity within and between hemispheres during motor imagery tasks, introducing network-based metrics such as integration and segregation. Evaluating these metrics across multiple EEG-based BCI datasets, our findings reveal that network lateralization offers neurophysiological plausible insights, characterized by stronger lateralization in sensorimotor and frontal areas contralateral to imagined movements. While these lateralization features did not outperform CSP and Riemannian geometry in terms of classification accuracy, they demonstrated competitive performance against PSD alone and provided biologically relevant interpretation. This study underscores the potential of brain network lateralization as a new feature to be integrated in motor imagery-based BCIs for enhancing the interpretability of noninvasive applications.
\end{small}

\vspace{10pt}
\noindent{\it Keywords\/}: Brain-Computer Interface, Feature interpretability, Brain Network, Common Spatial Pattern, Riemannian geometry.



\section{Introduction}

Brain-computer interfaces (BCIs) translate brain activity patterns into commands or messages for interactive applications \cite{vidal1973toward,bozinovski1988using,wolpaw2002brain}. These systems are increasingly being investigated for control and communication \cite{wolpaw2002brain,carmena2003learning,carlson2013brain}, as well as for restoring lost neurological functions caused by stroke or other nervous system injuries \cite{daly2008brain,vansteensel2016fully,kruse2020effect}. Many BCI applications rely on the ability of subjects to voluntarily modulate their brain activity through mental imagery. A prominent paradigm in this domain is motor imagery (MI), which relies on the imagination of kinesthetic movements of large body parts, engaging motor representations similar to actual movement execution 
\cite{jeannerod1995mental,pfurtscheller2001motor,lotze2006motor,guillot2009brain}. MI induces modulations in brain activity that lead to detectable signal changes, such as Event-Related Desynchronization or Synchronization (ERD/ERS) \cite{pfurtscheller1999event}. These changes manifest as specific amplitude variations in signal power within defined frequency bands, making power spectral density (PSD) the most conventional feature used to characterize MI.

Nonetheless, identifying mental intentions from brain signals requires working across different domains: temporal, frequency, and spatial. Given that brain signals are often characterized by noisy measurements and low spatial resolution, it is imperative to employ methods that enhance the distinctive characteristics that define each mental task. During an MI task, brain activation is prominently localized in the sensorimotor cortex. Then a smart solution to enhance signal quality is the application of spatial filtering \cite{lotte2014tutorial}. These methods aim to highlight relevant information while minimizing the influence of surrounding neural activity. Over the past few decades, the most widely adopted spatial filtering technique in the BCI field has been Common Spatial Patterns (CSP) \cite{ramoser1999designing,pfurtscheller2001motor,blankertz2005berlin,blankertz2007optimizing}. This filter works as a data-driven dimension reduction method, extracting signal sources by maximizing the variance ratio between two conditions. This technique relies on the simultaneous diagonalization of two covariance matrices derived from the band-pass filtered signals of the two classes. 

Other techniques have focused on the classification block, such as Riemannian methods, which have achieved outstanding accuracies and thus gained significant importance in the field \cite{barachant2013classification, jayaram2018moabb}. These methods enable direct manipulation of signal covariance matrices by leveraging manifold topology \cite{yger2016riemannian, congedo2017riemannian}. The core idea behind these algorithms is to work with covariance matrices in the manifold of symmetric positive-definite (SPD) matrices, using them as features in a classifier that respects their intrinsic geometry.
Nevertheless, these methods face two major disadvantages: high computational complexity and a risk of overfitting in high-density systems \cite{rodrigues2017dimensionality, congedo2017closed}. Additionally, they suffer from lack of interpretability, a significant but often overlooked problem. Riemannian methods do not provide a direct way to determine which parts of a signal are being used for classification, and classifiers operating within the manifold do not address this issue \cite{barachant2010riemannian, barachant2011multiclass}.

Although these approaches exhibit high accuracy, a non-negligible portion of subjects (approximately $\sim$30\%) still show inefficient performance \cite{thompson2019critiquing}. Besides, these methods suffer from a lack direct interpretability, leaving open the possibility that artifacts might influence classification results. Notably, none of these methods focus on the feature extraction block, raising concerns about the reliance of their predictions solely on brain-derived features. To bridge this gap, an novel approach has emerged, proposing the decoding of mental states through functional connectivity (FC) \cite{cattai2021phase, corsi2022functional, cao2022effective}. This method assesses the complexity of neurophysiological processes by measuring information exchange among different brain areas. Using network theoretic techniques, these interactions are analyzed to extract key summary properties from the intricate brain network \cite{boccaletti2014structure, de2014graph, gonzalez2021network}. Moreover, by integrating network topology with the brain's spatial layout, this approach uncovers crucial insights into brain function, offering a powerful tool for understanding the intricacies of neural communication \cite{gotts2013two, liu2009evidence}. Multiple neuroimaging studies have demonstrated that during MI tasks, the brain exhibits a distinct spatial activation pattern, characterized by predominant activation of the contralateral hemisphere over the motor cortex \cite{beisteiner1995mental,pfurtscheller1999event,xu2014motor,cattai2021phase}. Building upon these findings, we investigated the dual contribution of brain network topology and space on modeling motor-related mental states through the concept of functional lateralization. Specifically, we introduced novel metrics to assess segregation and integration within and between hemispheres, demonstrating their significant relevance in decoding MI mental tasks.

\section{Materials and Methods}
\label{Materials and Methods}

\subsection{EEG Dataset cohorts}
We designed our experimental setup based on a open source benchmark MOABB (\textit{Mother of all BCI Benchmark}) \cite{jayaram2018moabb}. We selected nine open-access datasets of healthy participants, which contain non-invasive EEG signals recorded during MI experiments that focused on left and right hand grasping motions. 
\textbf{Table~\ref{tab:datasets}} provides a description of the selected datasets. Each trial was band-passed filtered in a broad $\alpha$-$\beta$ band ($8$-$35Hz$), where characteristic signal changes during MI tasks are typically observed \cite{pfurtscheller1977event, pfurtscheller1999event, neuper2001event}.

\subsection{Building Functional Brain Networks}
FC presumes interaction between distant brain areas if there is statistical synchronization between their activities. There are several FC estimators \cite{bastos2016tutorial}; in this study, we implemented undirected spectral coherence ($w$) \cite{carter1987coherence}, which is well-documented in the MI-BCI domain \cite{hamedi2016electroencephalographic, cattai2021phase, corsi2022functional}. This estimator is constructed by computing the normalized cross-spectral density between signals from two electrodes $i$ and $j$ ($i\neq j$ and $i,j \in \{1,2,\ldots,N \}$), for a particular frequency $f$:

\begin{equation}
w_{ij}[f] =\frac{|P_{ij}[f]|}{(P_{i}[f]P_{j}[f])^{1/2}}, 
\label{Eq:w}
\end{equation}

where $P_{ij}[f]$ represents the cross-spectrum and $P_{i}[f]$ the auto-spectrum. These FC values are integrated within the $\alpha$-$\beta$ band (8-35 Hz).

We estimated the cross-spectral density of each pair of EEG signals at the trial level using multitapers \cite{slepian1978prolate} with 1-second time windows and 0.5-second overlap, with a frequency resolution of 1 Hz. We averaged the resulting FC matrices over $\alpha$-$\beta$ bands. Consequently, for each trial, we obtained a $W$ symmetric adjacency matrix of shape $N$ × $N$, where $N$ is equivalent to the number of EEG channels. These matrices correspond to fully connected and weighted networks.

\subsubsection{Spatial network lateralization metrics}
\begin{figure}[!t]
\begin{center}
\includegraphics[width=1\linewidth]{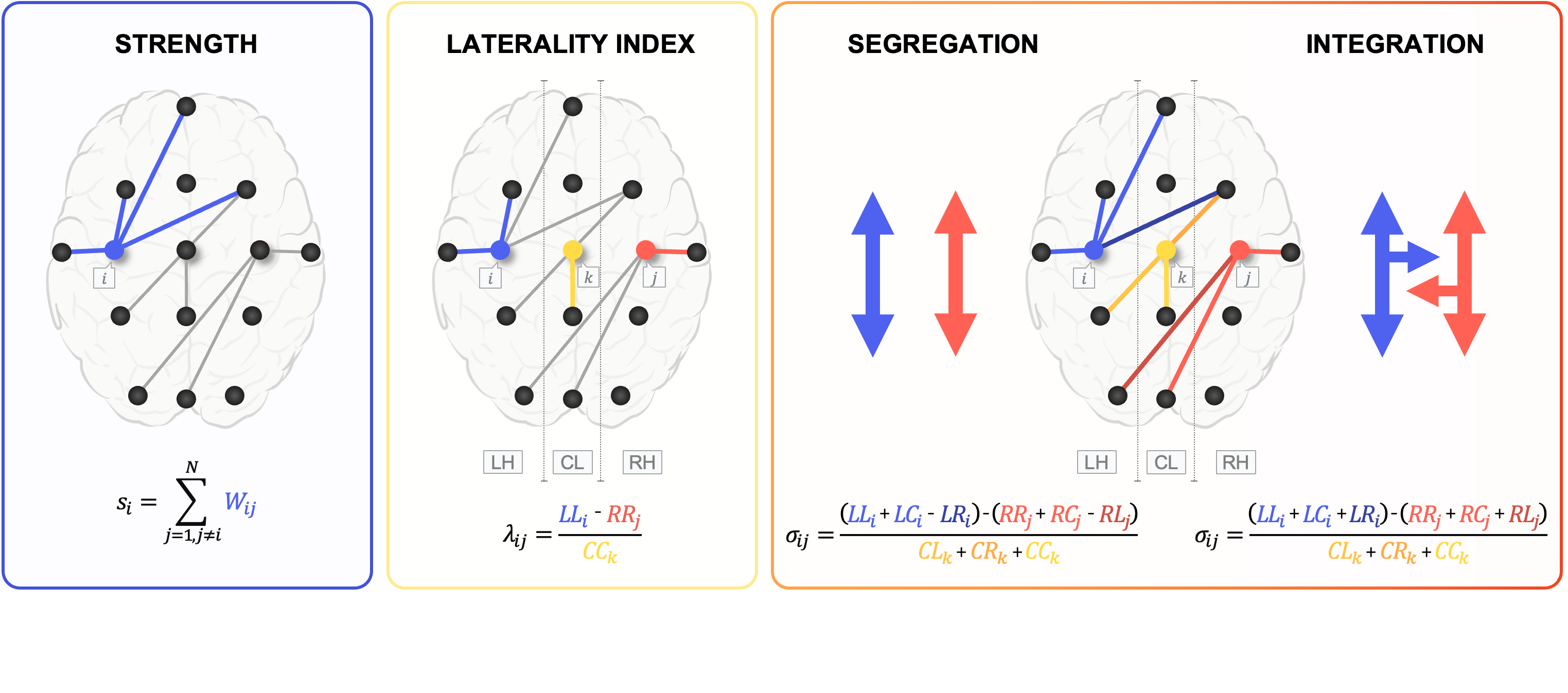}
    \caption[Network properties.]{\textbf{Network properties}. Functional lateralized nodes can be identified by comparing the strength between homotopic pairs. These three figures illustrate how links within the same network are used to compute lateralization properties. 
    The top left figure represents the computation of the $strength$ of node $i$. The top right, introduced $laterality$ $index$ for the homotopic pair $i$-$j$ ($\lambda_{ij}$).     The bottom figure represents the distinction between $segregation$ ($\sigma$) and $integration$ ($\omega$) at the same pair $ij$. 
    The key difference lies in the influence of inter-hemispheric links ($LR_{i}$ and $RL_{j}$). While $\omega$ aggregates the strength of bilateral interactions, $\sigma$ measures the strength of within-hemisphere interactions. A large positive $\sigma$ suggests a stronger bias for within-hemisphere interactions in the left hemisphere, whereas a large negative value indicates a stronger bias in the right hemisphere. The notations follow the conventions outlined in \textbf{Box~\ref{box1}}. LH: left hemisphere, CL: center line, RH: right hemisphere}.
    \label{fig:LatMet}
\end{center}
\end{figure}

{\label{LatMethod}}
In weighted networks, edges can assume a range of values, reflecting a hierarchy of connections that yield varied levels of connectivity for each node. This phenomena is perfectly captured by node strength ($s$), a fundamental property that serves as a reference for studying connection patterns related to brain lateralization (\textbf{Figure~\ref{fig:LatMet}}). If we denote $W$ as the weighted connectivity matrix of the $N$-nodes brain network, then we can compute the $s$ of node $i$ as 

\begin{equation}
s_{i}=\sum_{j=1, j{\ne}i}^{N}W_{ij},
\label{Eq:S}
\end{equation}

To explore lateralization, we focused on pairs of homotopic nodes, which means mirror channels across the hemispheres. For instance, in the 10-20 international system EEG configuration, nodes C3 and C4 are defined as homotopic. Then we can estimate the laterality index of the homotopic pair $i$ and $j$ ($\lambda_{ij}$) by measuring the intra-hemisphere strength difference between them, normalized by the strength of the closest midline node $k$:

\begin{equation}
\lambda_{ij} =\frac{LL_{i} - RR_{j}}{CC_{k}}.
\label{Eq:lat}
\end{equation}

For a clarified notation, check \textbf{Box~\ref{box1}}. In \textbf{Figure~\ref{fig:LatMet}} we illustrate how these interactions are distributed for a toy example network. 

The concept of functional lateralization can be further developed by analyzing the influence of interactions within and across hemispheres. By adapting the metrics proposed by \cite{gotts2013two} and \cite{liu2009evidence}, segregation can be defined as the tendency for greater within-hemisphere interactions compared to between-hemisphere interactions. This is calculated as the difference between intra- and inter-hemispheric strength (i.e., $LL_{i} + LC_{i} - LR_{i}$ or $RR_{j} + RC_{j} - RL_{j}$) (\textbf{Figure~\ref{fig:LatMet}}). Specifically, the lateralization of segregation for a pair of homotopic nodes $i$ and $j$ is determined by calculating their segregation difference:

\begin{equation}
\sigma_{ij} =\frac{(LL_{i} + LC_{i} - LR_{i}) - (RR_{j} + RC_{j} - RL_{j})}{(CL_{k} + CR_{k} + CC_{k})} 
\label{Eq:seg}
\end{equation}

The analysis of this metric's sign could lead to some misinterpretation, so it is necessary to emphasize two main aspects. First, the strength values involved in the equation are strictly positive since we are working with undirected networks. Second, to guarantee a true sided $\sigma_{ij}$, we empirically proved that $LL_{i} + LC_{i} > LR_{i}$ and $RR_{j} + RC_{j} > RL_{j}$ for every node (see \nameref{Appendix} \textbf{Figure~\ref{fig:t-test_LC_RC}}). This means that a negative $\sigma_{ij}$ value reflects higher lateralization of segregation in the right homotopic node of the pair. In other words, within-hemisphere interactions are stronger in the right hemisphere. The opposite situation occurs for a positive value.

On the other hand, integration ($\omega_{ij}$) seeks the contribution of inter-hemispheric connections, characterizing how information flows across hemispheres. In the mathematical formulation, it is translated as the summed effect of intra- and inter-hemispheric interactions (e.g., $LL_{i} + LC_{i} + LR_{i}$ or $RR_{j} + RC_{j} + RL_{j}$) (\textbf{Figure~\ref{fig:LatMet}}). Therefore, the lateralization of integration for a node $i$ in the left hemisphere compared to node $j$ in the right hemisphere is calculated as:

\begin{equation}
\omega_{ij} =\frac{(LL_{i} + LC_{i} + LR_{i}) - (RR_{j} + RC_{j} + RL_{j})}{(CL_{k} + CR_{k} + CC_{k})} 
\label{Eq:intg}
\end{equation}

As a general remark, it is important to highlight that all these properties are local, meaning they characterize each node individually. From a classification perspective, this implies that the number of nodes is equivalent to the number of features. However, the lateralization metrics reduce the number of features to half minus the number of nodes in the central line, as each pair of homotopic nodes has the same feature value but with opposite signs (e.g., $\lambda_{ij} = - \lambda_{ji}$).

\subsubsection{Statistical analysis}
\label{Meth_stats}
Given that hand-MI tasks are known to manifest as lateralized activities in the motor cortex, we hypothesized that the proposed lateralization metrics hold significant potential for distinguishing between left and right hand-MI. To statistically assess this capability, we conducted a 5000-permutation $t$-test for each metric. Comparisons were made at the subject level, with the number of samples equivalent to the number of trials. A significance level of $p<0.05$ was considered critical for determining statistical significance. By performing this analysis for each node, we were able to identify the most discriminative electrodes. This test was repeated for the entire subject population.

\subsubsection{Feature Selection}
Since the number of features increases proportionally with the number of nodes, there is a risk of overfitting, especially with datasets containing a large number of channels. To mitigate this issue, we implemented a feature selection step to limit the number of features and reduce the classifier's parameter load. It is important to note that feature selection does not entail node removal. Instead, the interactions of non-selected nodes with the selected ones continue to contribute to the latter's representation.

The feature selection algorithm employed an embedded approach to identify the most discriminant features. First, we normalized the features by applying a z-score transformation. Then, we used sequential forward feature selection within a nested 5-fold cross-validation (CV) framework using linear kernel Support Vector Machine (SVM). This algorithm progressively adds features to build a feature subset. To streamline the process, features were ranked based on their discrimination power using a $t$-test within each training fold. Subsequently, a bottom-up search procedure was performed, conditionally including new features into the selected set based on the inner-CV score. This process continued until the score ceased to increase or until the maximum number of features, equivalent to the number of samples, was reached. Finally, classification performance was measured in terms of ROC-AUC. The code is publicly available at: \url{https://github.com/julianagonzalezastudillo/netfeat}

\subsection{Power Spectrum Density}
To establish a well-defined benchmark for comparing network-based features, we computed the power spectrum density (PSD) using Welch's method with a Hamming window and 50\% overlap. The PSD estimates were averaged over the $\alpha$ and $\beta$ bands to capture the frequency-specific power variations associated with MI. This provided a reference feature set for our analysis. For consistency, these features were also z-score transformed, and the same feature selection algorithm used for network features was also applied.

\subsection{Common Spatial Pattern}
CSP generates $N$ spatial filters, being $N$ equivalent to the number of electrodes. Yet, it is necessary to select an optimal subset of components to capture the difference between classes while avoiding overfitting. Here, we utilized eight components. Subsequently, the original signals were projected onto these selected spatial filters, and the logarithmic power was computed. Finally, the resulting eight-dimensional log-variances were linearly combined to serve as features for a linear SVM classifier \cite{lotte2007review}.

Additionally, from the spatial filter decomposition, we extracted the corresponding patterns of brain activation by taking the inverse of the transposed full filters matrix \cite{blankertz2005berlin, blankertz2007optimizing, barachant2010common}. These spatial patterns represent the projected sources on the scalp and can be used to validate the neurophysiological likelihood of the extracted features.

\subsection{Riemannian space}
Despite the lack of direct neurophysiological interpretability of Riemannian methods, some solutions have been proposed \cite{barachant2011channel, xu2020tangent, larzabal2021riemannian, qu2022riemannian}. Here, we implemented the Riemannian-based feature selection introduced in \cite{barachant2011channel}. This algorithm uses the Riemannian distance between the class-conditional mean covariance matrices as the selection criterion in a backward selection approach. In an iterative loop, we retain the top $N*$ electrodes that maximize the criterion, where $N*$ is a predefined value lower than $N$ (set to 10 for our analysis). It's worth noting that each electrode represents the $i$-th row and column in the covariance matrix.
To ensure consistency in the classification approach across methods, we vectorized the reduced covariance matrices by mapping them onto the tangent space of the Riemannian manifold at the geometric mean of the set of covariance matrices \cite{barachant2011multiclass}.

\section{Results}
{\label{Results}}

To comprehensively assess the performance of the proposed lateralization properties, three aspects were considered in this paper: their statistical differentiation power between MI conditions, their neurophysiological plausibility and their classification performance.

\subsection{Classification performance}
\begin{figure}[h!]
\begin{center}
\includegraphics[width=1\linewidth]{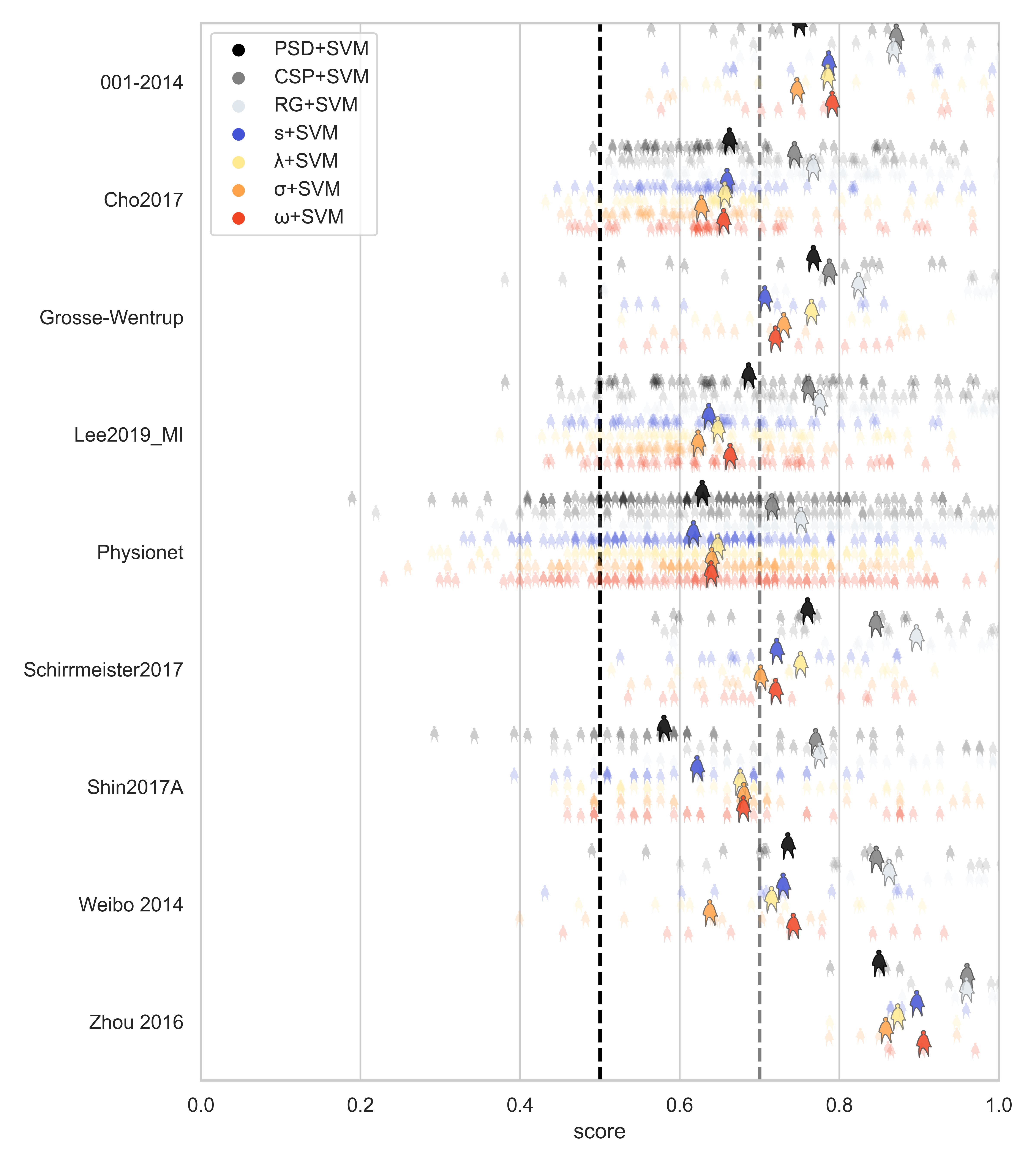}
    \caption[Classification performances.]{\textbf{Classification performances.} Classification scores for each method were evaluated across datasets using a 5-fold CV SVM. 
    Each feature extraction method follows a specific process to prepare a proper input for the classifier. Typical PSD is computed using Welch's method, followed by a sequential feature selection to avoid overfitting. Network features undergo a nested-CV selection to reduce dimensionality and ensure the most discriminant nodes. The CSP method projects the signal using selected spatial filters and then computes the logarithm of the power of the projected signal. Lastly, reduced-Riemannian SPD matrices are projected and vectorized on the tangent space of the manifold. All types of features converge in separate SVM classifiers.
    Each transparent silhouette represents a single subject, while the larger contoured silhouette represents the mean across subjects. Note that each subject has only one score, representing the mean between sessions (if applicable, see \textbf{Table~\ref{tab:datasets}}). The black dotted line indicates chance level performance (0.5), and the grey line marks the threshold for efficient performance (0.7) \cite{thompson2019critiquing}.}
    \label{fig:classif_acc}
\end{center}
\end{figure}

\textbf{Table~\ref{tab:classif_acc}} and \textbf{Figure~\ref{fig:classif_acc}} present the results generated by the entire processing chain described in \nameref{Materials and Methods}. The group-averaged scores across datasets were quite heterogeneous, independent of the pipeline used. Variations in hardware, strategy paradigms, and individual subjects contribute to significant differences in BCI task outcomes, making it challenging to generalize findings. Despite this, we observed that CSP and Riemannian methods consistently outperformed PSD and network-based features, with Riemannian geometry achieving the highest accuracies. 

\begin{figure}[!htb]
\begin{center}
\includegraphics[width=1\linewidth]{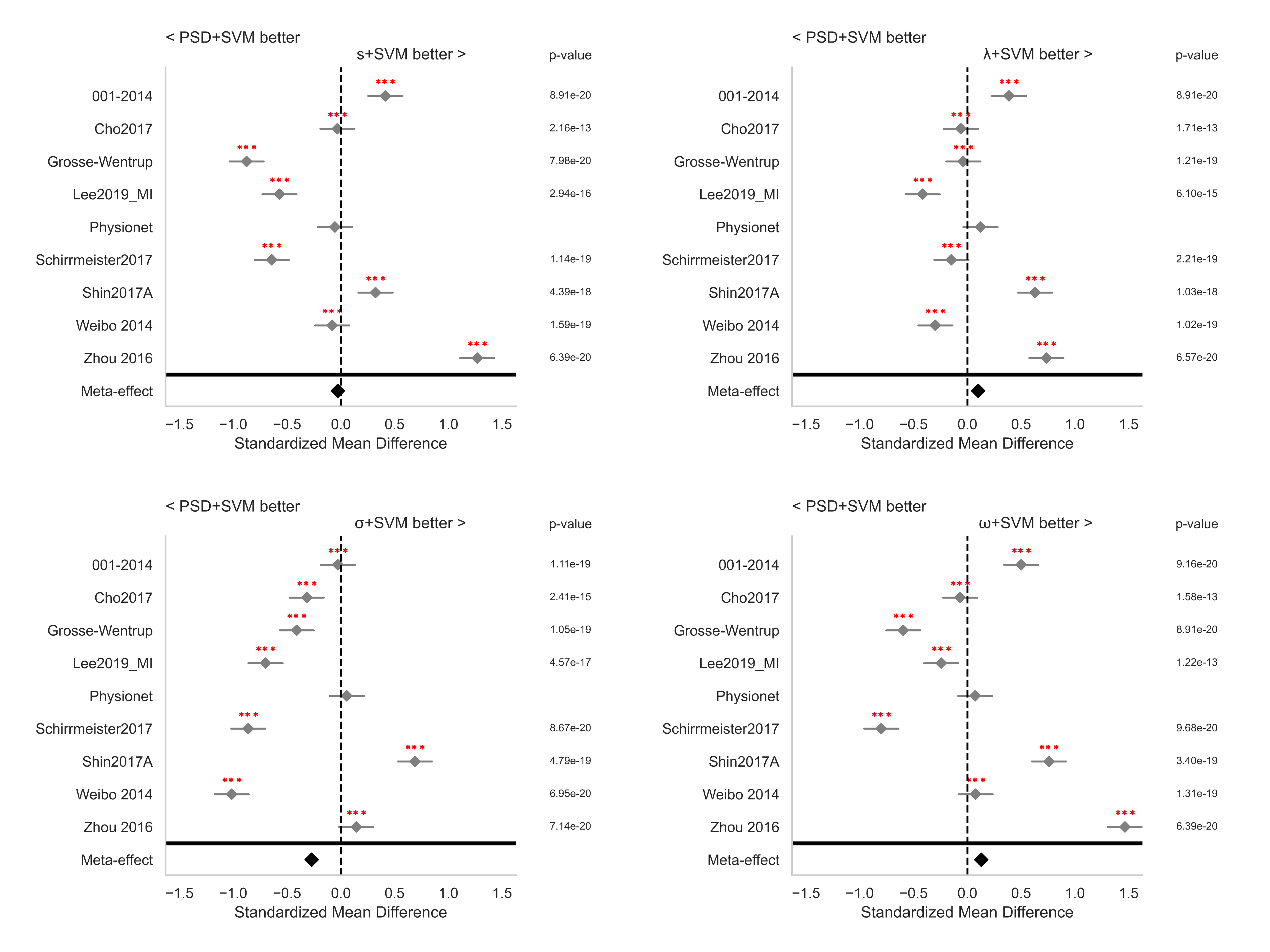}
    \caption[Classification statistical analysis: network properties versus PSD.]{\textbf{Classification statistical analysis: network properties versus PSD.} Meta-analysis-style plots depict the performance comparison between lateralization network metrics and PSD. The effect sizes displayed are standardized mean differences, ith $p$-values corresponding to the one-tailed Wilcoxon signed-rank test for the hypothesis given at the top of the plot and 95\% interval denoted by the grey bars. Significance levels are indicated by stars: *** = $p < 0.001$, ** = $p < 0.01$, * = $p < 0.05$. The meta-effect, displayed at the bottom of the plots, underscores that laterality and integration outperform PSD.}
    \label{fig:classif_stats}
\end{center}
\end{figure}

Nonetheless, network-based features demonstrated performance comparable to the conventional PSD method. The meta-analysis indicated no clear tendency for one method to consistently outperform the other, exhibiting a non-significant meta-effect (\textbf{Figure~\ref{fig:classif_stats}}). Examining the details, there was an overall trend across datasets: \textit{Cho2017}, \textit{Grosse-Wentrup}, \textit{Lee2019\_MI}, and \textit{Schirrmeister2017} consistently performed better with PSD, while \textit{Shin2017A} and \textit{Zhou2016} showed better results with network features. Interestingly, when processed with the PSD pipeline, the \textit{Shin2017A} dataset obtained the lowest score of all, revealing a possible trend for better results with network features for subjects who achieve low scores with PSD.

\subsection{Features interpretation}
\subsubsection{Network lateralization}
Regardless of the superiority of CSP and Riemannian methods in terms of classification performance, it is essential to prove which are the underlying physiological mechanisms that drive these results. For network properties, one approach is to identify the nodes that best differentiate between MI states. To this end, we applied the statistical analysis described in materials and methods. For simplicity and to avoid any confusion between MI classes and hemisphere sides, we refer to left-hand MI as $LMI$ and right-hand MI as $RMI$.

\begin{figure}[h!]
\begin{center}
\includegraphics[width=1\linewidth]{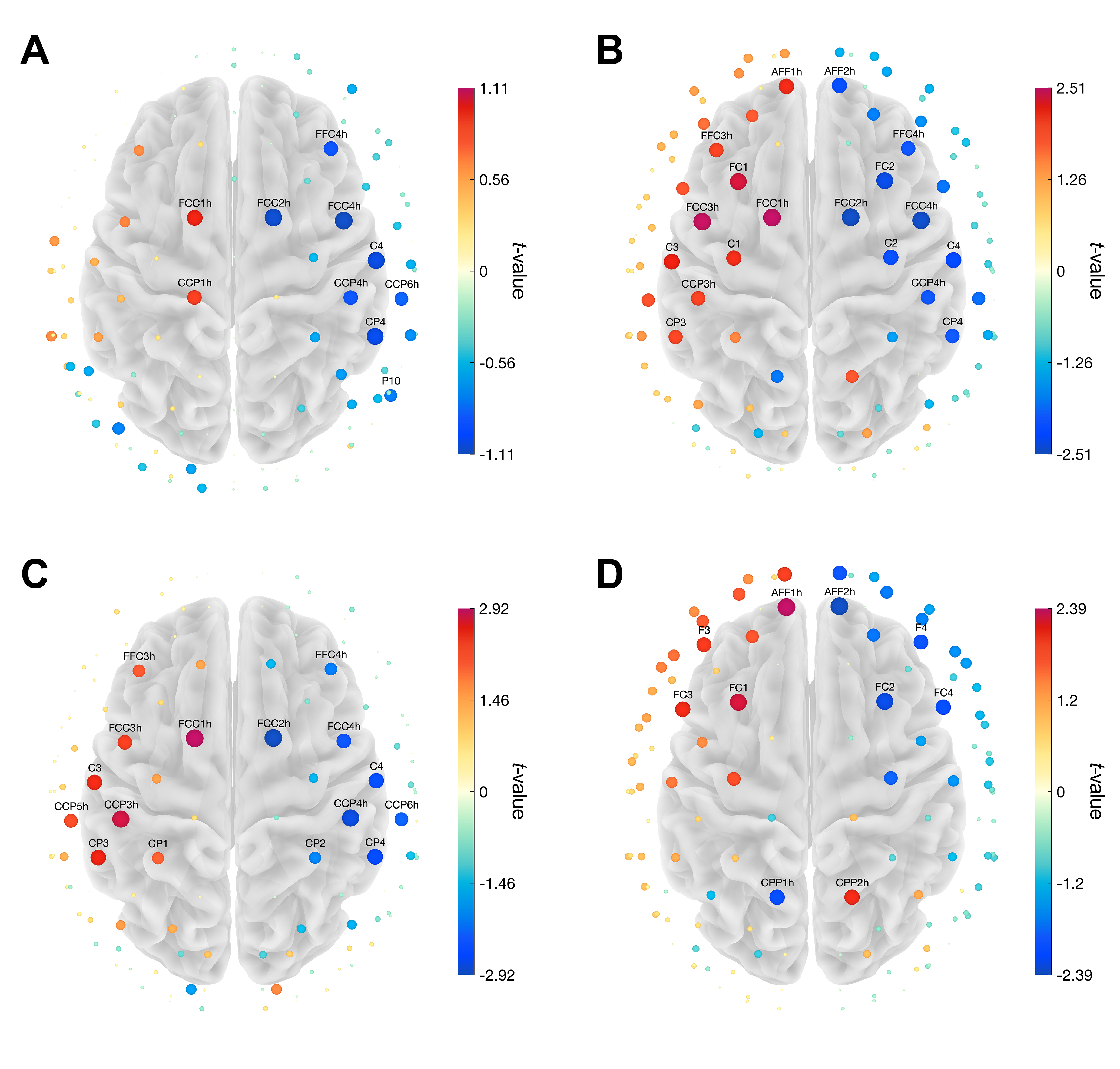}
    \caption[Network features in MI tasks.]{\textbf{Network features in MI tasks.}
    Group-averaged $t$-values, contrasting $RMI$ versus $LMI$ in the $\alpha$-$\beta$ band for the four network properties under study. All lateralization metrics are inversely hemisphere-symmetrical.
    \textbf{A. Strength:} evidence of hemisphere lateralization is observed in motor-related areas, with a predominance of higher values in the right hemisphere. For illustrative purposes, only the names of the ten nodes with the highest values are shown.
    \textbf{B. Laterality Index:} this metric accentuates the differences between the two MI tasks, showing nine significant nodes ($p<0.05$) in the posterior frontal cortex, precentral and postcentral gyrus and superior parietal cortex. Only the names of nodes with significant $t$-values are displayed.
    \textbf{C. Integration:} seven significant nodes are mostly located over the postcentral gyrus and superior parietal cortex, principally in somatosensory areas.
    \textbf{D. Segregation:} four significant nodes are observed, with a tendency for higher values in the posterior frontal cortex and dorsolateral prefrontal cortex.
    Note that the \textit{Grosse-Wentrup} dataset is excluded from feature analysis due to a different electrode naming convention.  
    }
    \label{fig:t-test_net}
\end{center}
\end{figure}

\begin{sidewaysfigure}[h!]
\begin{center}
\includegraphics[width=1\linewidth]{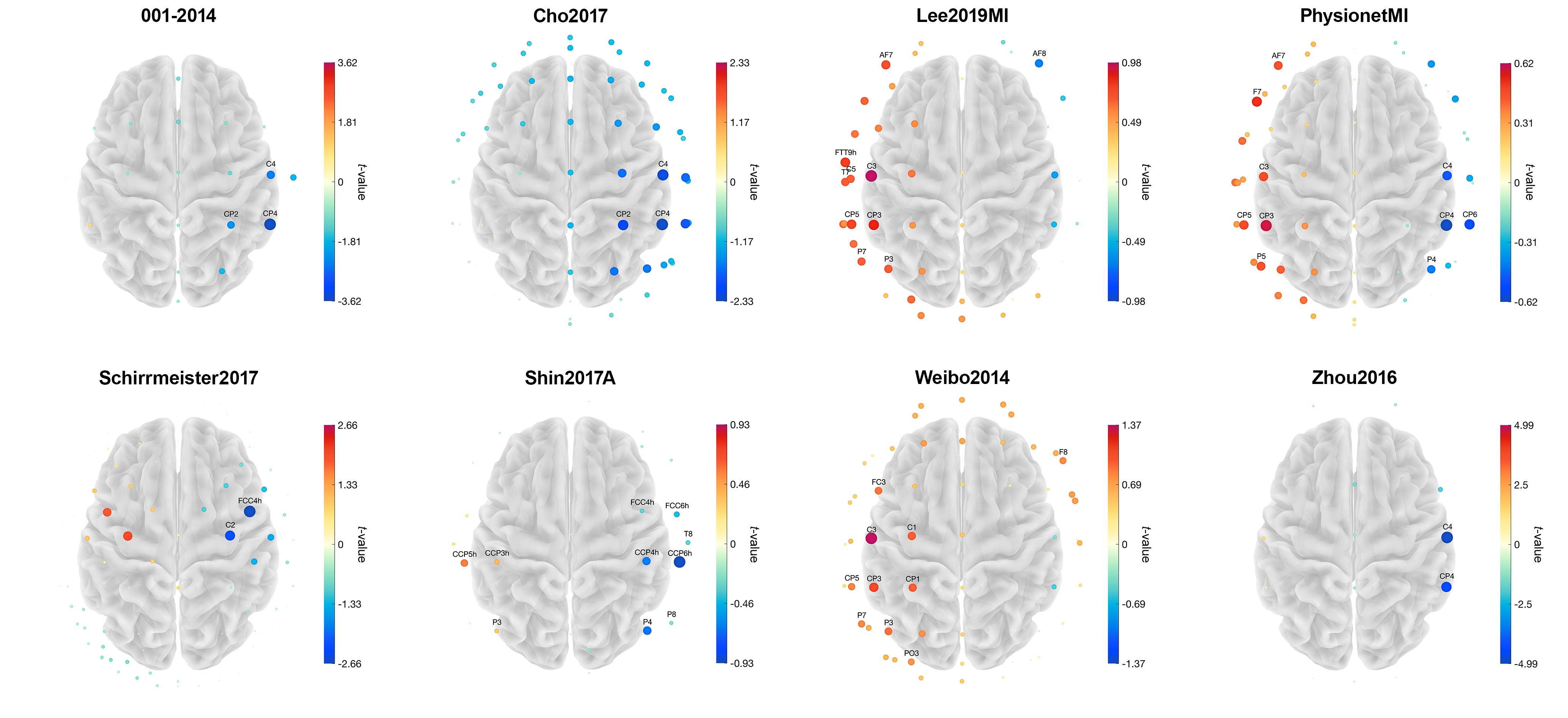}
    \caption[Strength.]{\textbf{Strength.} Group-averaged $t$-values contrasting strength for $RMI$ versus $LMI$ in the $\alpha$-$\beta$ band for each dataset.}
    \label{fig:sup_t_test_strength_all_dt}
\end{center}
\end{sidewaysfigure}

In \textbf{Figure~\ref{fig:t-test_net}A} we show the node strength $t$-values obtained across trials and averaged across subjects. These results revealed interesting patterns for a subset of nodes. Notably, the largest changes tend to concentrate in motor-related areas. Even more striking was the predominance of positive $t$-values in the left hemisphere. This trend confirms that $RMI$ evokes higher strength in the contralateral motor cortex. The inverse situation occurred for $LMI$ with even stronger $t$-values and more recruited areas, suggesting that this task requires more connectivity resources. The same pattern was observed for each individual dataset \textbf{Figure~\ref{fig:sup_t_test_strength_all_dt}}. 

This evidence of sided-contrast connections across tasks encouraged further lateralization analysis. For this purpose, the introduced network metrics consider the spatial locations of the electrodes, distinguishing between intra- and inter-hemispheric interactions. These metrics yield symmetric inverse values for each pair of homotopic nodes. Also notice that the sign of the $t$-values is strictly related to the task, i.e. a positive value means stronger lateralization for $RMI$ and a negative for $LMI$.

\textbf{Laterality index}.
When applying the same statistical analysis to lateralization metrics, we observed comparable behavior between strength and $\lambda$ (\textbf{Figure~\ref{fig:t-test_net}B}). Notably, the $t$-values were accentuated, suggesting that combining homotopic information enhances the differentiation between MI tasks. The highest $t$-values were predominantly located in MI-related areas: the dorsolateral prefrontal cortex (DLPFC), premotor cortex (PMA), supplementary motor area (SMA), primary motor cortex (M1), primary somatosensory cortex (S1), and somatosensory association cortex (S2) \cite{jeannerod1999mental,grezes2001functional,guillot2009brain,hetu2013neural}.

\textbf{Integration}.
In the same line, considering the contribution of inter-hemispheric interactions, $\omega$ increased the difference over nodes related to motor planning (PMA and SMA) and execution (M1), as well as integrating sensory information (S1 and S2) (see \textbf{Figure~\ref{fig:t-test_net}C}). Distinctively, this metric maintained significant $t$-values for channels associated with MI while reducing the rest.

\textbf{Segregation}.
For the specific case of $\sigma$, we first queried whether the connections with the middle line nodes ($LC_{i}$ and $RC_{j}$) should be considered within-hemisphere. We analyzed their role in the two possible scenarios by statistically comparing the differences. From the behaviors shown in \textbf{Figure~\ref{fig:t-test_LC_RC}}, we concluded that middle line connections more appropriately belong to within-hemisphere. This conclusion is drawn from the fact that nodes with predominant connections of this type are closer to the middle line, and many are strategic for the MI tasks under study. Then, reducing their influence by subtracting $LC_{i}$ and $RC_{j}$ links may alter the neurophysiological nature of the results. Additionally, this helps avoid any misinterpretation of the sign of $\sigma$.

When analyzing the impact of subtracting the now well-defined inter-hemispheric connections ($LR_{i}$, $RL_{j}$), $\sigma$ showed the highest impact in the frontal-central electrodes (\textbf{Figure~\ref{fig:t-test_net}D}). These nodes are primarily linked with SMA and PMA cortex, along with the dorsolateral prefrontal cortex (DLPFC) associated with action planning \cite{jeannerod1999mental,curtis2003persistent,gao2011evaluation,mokienko2013increased,hetu2013neural}. \newline

\begin{sidewaysfigure}[h!]
\begin{center}
\includegraphics[width=1\linewidth]{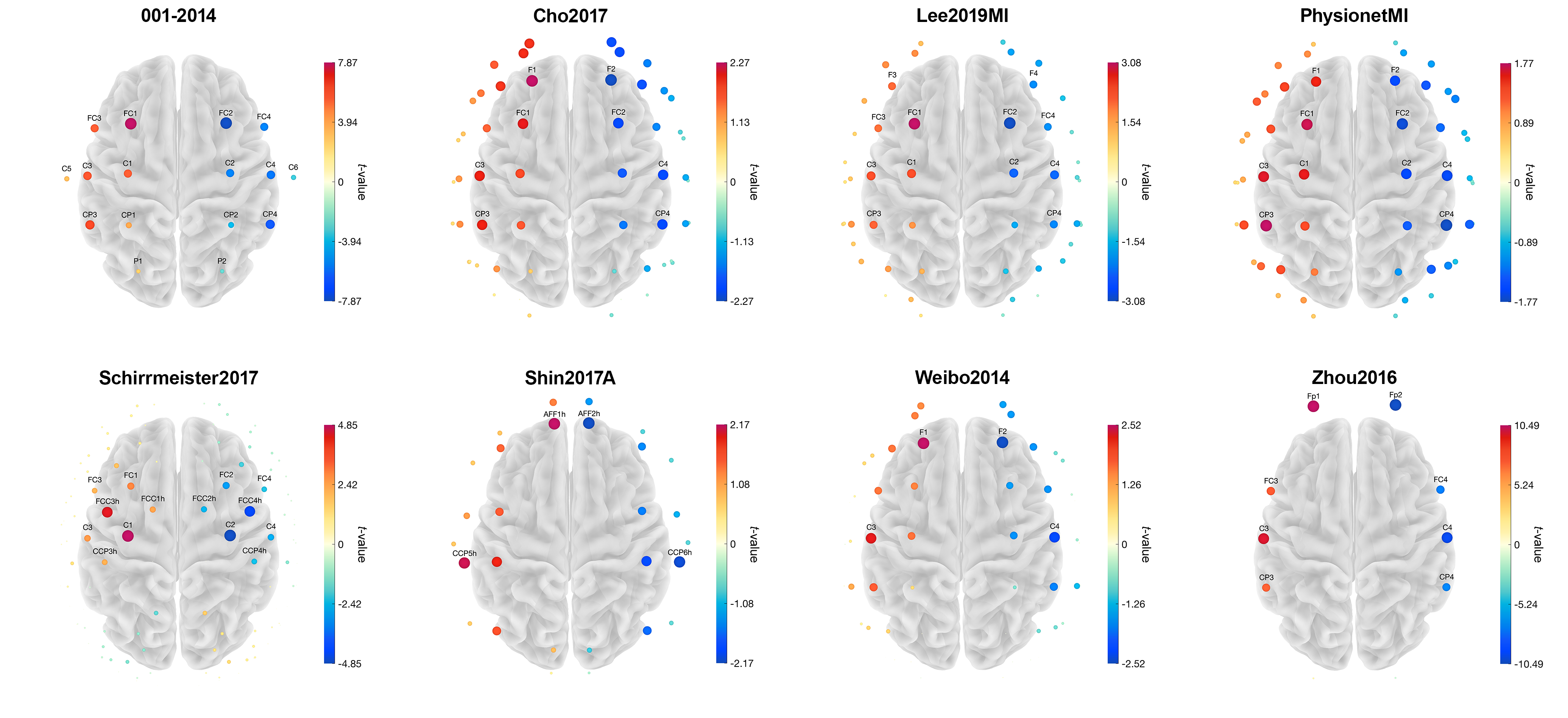}
    \caption[Laterality index.]{\textbf{Laterality index.} Group-averaged $t$-values contrasting laterality index for $RMI$ versus $LMI$ in the $\alpha$-$\beta$ band for each dataset.}
    \label{fig:sup_t_test_local_laterality_all_dt}
\end{center}
\end{sidewaysfigure}

\begin{sidewaysfigure}[h!]
\begin{center}
\includegraphics[width=1\linewidth]{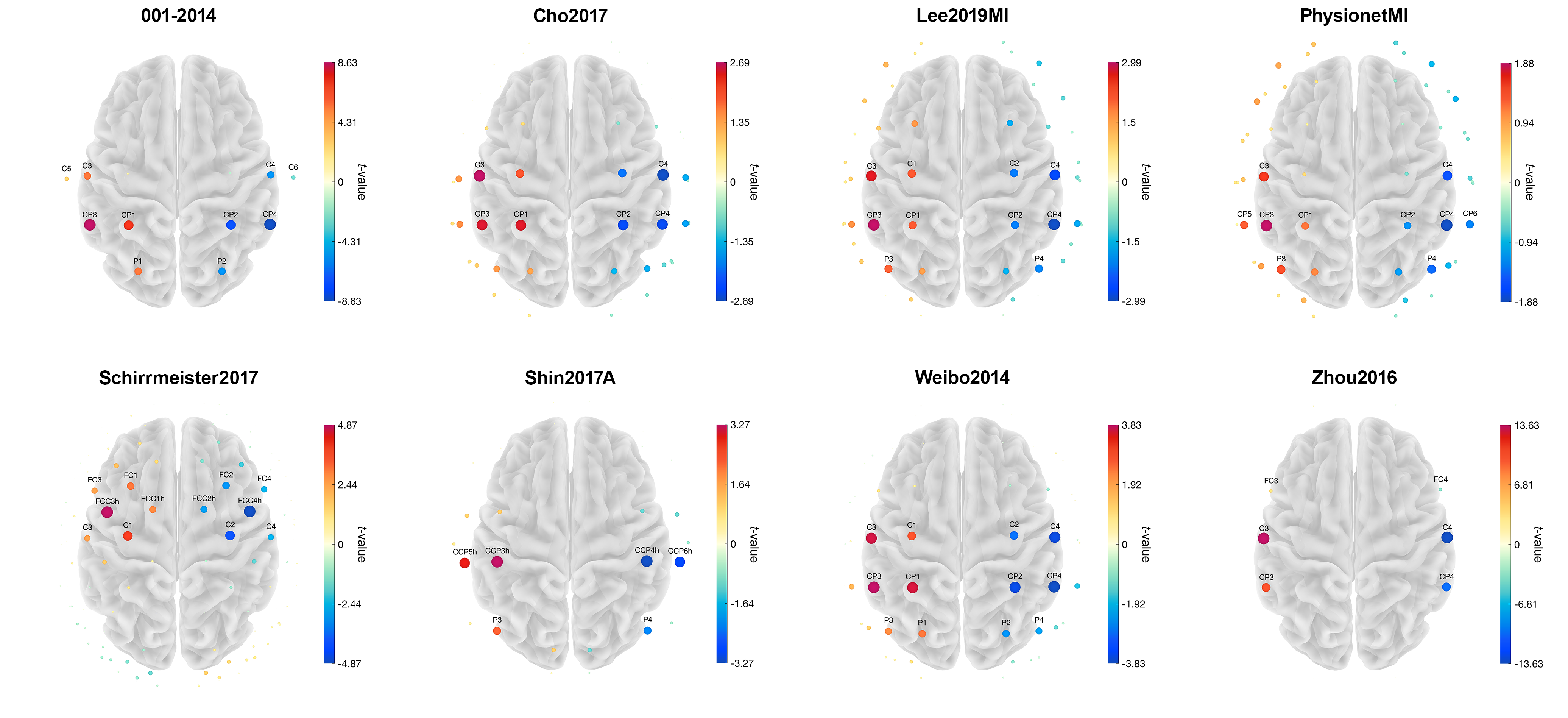}
    \caption[Integration.]{\textbf{Integration.} Group-averaged $t$-values contrasting integration for $RMI$ versus $LMI$ in the $\alpha$-$\beta$ band for each dataset.}
    \label{fig:sup_t_test_integration_all_dt}
\end{center}
\end{sidewaysfigure}

\begin{sidewaysfigure}[h!]
\begin{center}
\includegraphics[width=1\linewidth]{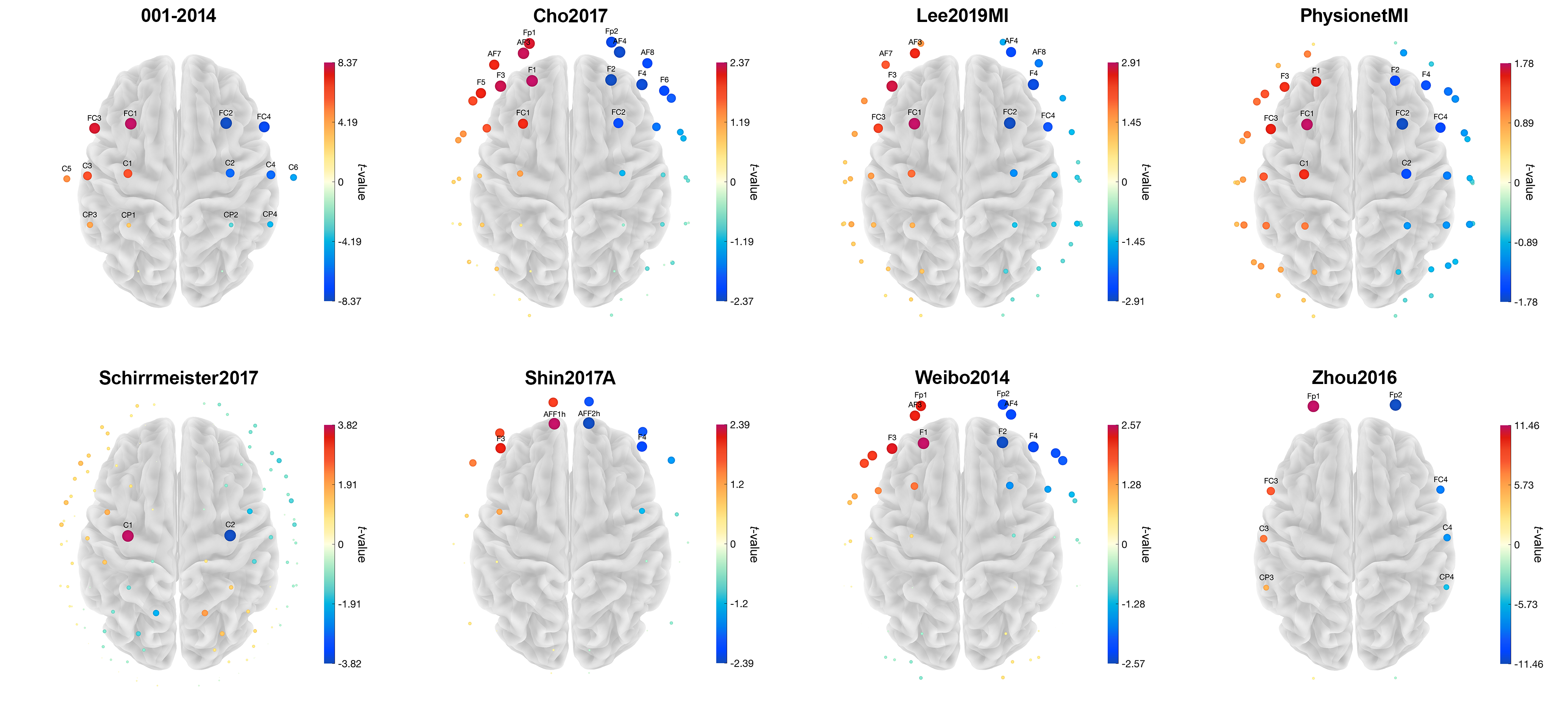}
    \caption[Segregation.]{\textbf{Segregation.} Group-averaged $t$-values contrasting segregation for $RMI$ versus $LMI$ in the $\alpha$-$\beta$ band for each dataset.}
    \label{fig:sup_t_test_segregation_all_dt}
\end{center}
\end{sidewaysfigure}

These findings suggest that MI of the hand grasping elicits detectable brain network changes, potentially aiding to characterize and discriminate MI-based BCI tasks. These changes revealed two simultaneous patterns of lateralization (i.e. fronto-central $\sigma$, while central-parietal $\omega$), primarily implicating sensorimotor areas. The same pattern was reproduced for each individual dataset (\textbf{Figure~\ref{fig:sup_t_test_local_laterality_all_dt}}, \textbf{Figure~\ref{fig:sup_t_test_integration_all_dt}}, \textbf{Figure~\ref{fig:sup_t_test_segregation_all_dt}}).

\subsubsection{Power Spectrum Density}
\begin{sidewaysfigure}[h!]
\begin{center}
\includegraphics[width=1\linewidth]{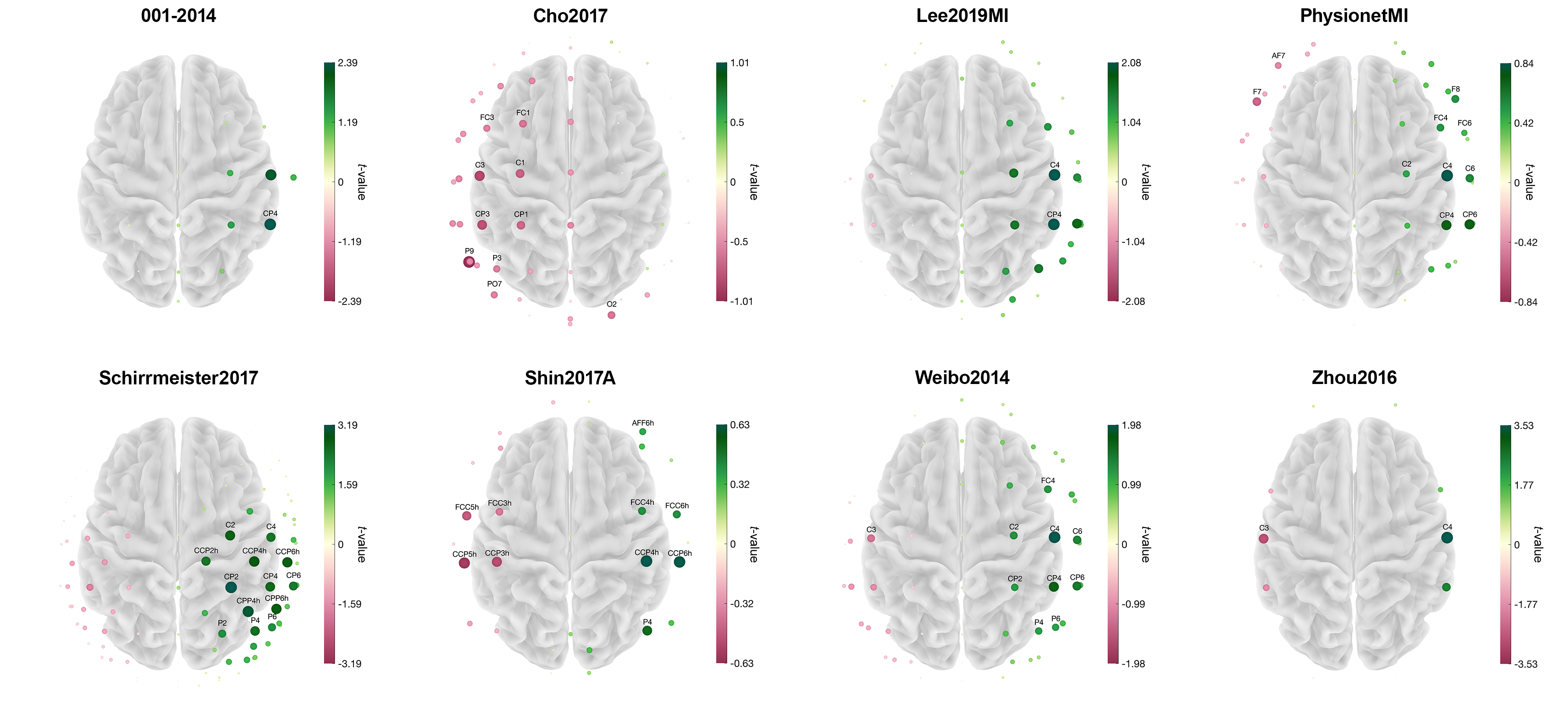}
    \caption[PSD.]{\textbf{PSD.} Group-averaged $t$-values contrasting PSD for $RMI$ versus $LMI$ in the $\alpha$-$\beta$ band for each dataset.}
    \label{fig:sup_t_test_psdwelch_all_dt}
\end{center}
\end{sidewaysfigure}

Observing the parity in classification performance between PSD and network-based features raised the question of whether the features driving PSD performance had significant neurophysiological plausibility. To explore this, we applied the same statistical analysis to PSD features. The results presented in \textbf{Figure~\ref{fig:state_methods}A}, showed significant mean $t$-values for only three electrodes, all located over the S1 and S2 areas of the right hemisphere. Overall, there was a tendency for enhanced values over the parietal cortex, predominantly on the right hemisphere. This pattern was consistent across almost all datasets, with the best differentiating electrodes observed in the right C and CP lines (see \textbf{Figure~\ref{fig:sup_t_test_psdwelch_all_dt}}). The exception was the \textit{Cho2017} dataset, which exhibited a tendency for more discriminating values on the left hemisphere. When comparing the discrimination power of PSD with network lateralization at the dataset level, lateralization properties consistently demonstrated higher discrimination power than PSD across each dataset.

\subsubsection{Common Spatial Pattern}
\begin{figure}[!htb]
\begin{center}
\includegraphics[width=1\linewidth]{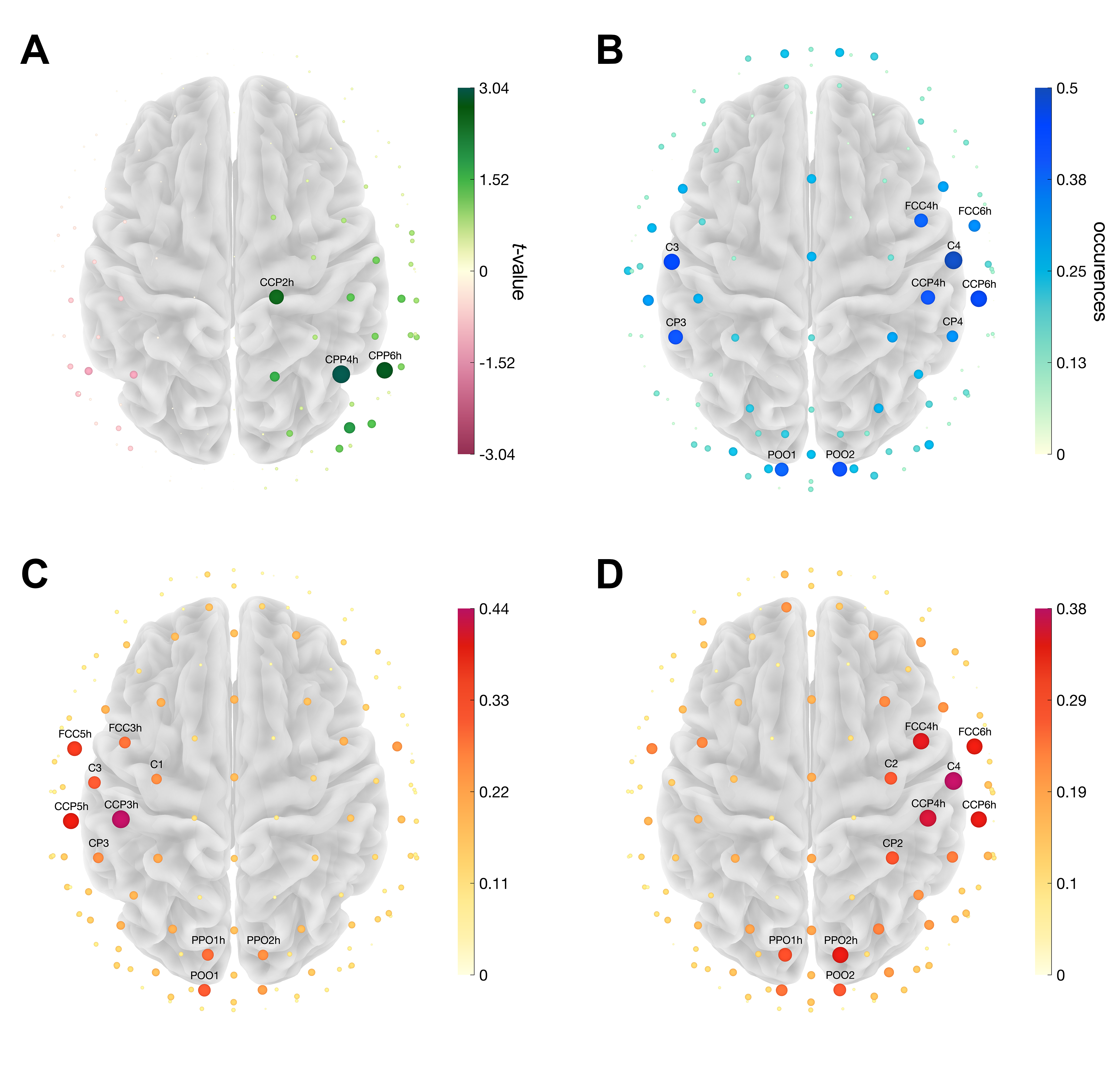}
    \caption[Feature analysis for state-of-the-art methods.]{\textbf{Feature analysis for state-of-the-art methods.} 
    \textbf{A. PSD:} group-averaged $t$-values contrasting $RMI$ versus $LMI$ in the $\alpha$-$\beta$ band. A positive value indicates spectral attenuation for $LMI$ and vice versa. Only nodes exhibiting significant values ($p<0.05$) are shown.
    \textbf{B. Riemannian occurrences:} group-averaged normalized occurrences, showing the number of times a specific feature in the manifold has been chosen. Most selected electrodes are located over the right and left motor-related cortex, as well as the occipital cortex.
    \textbf{C, D. CSP filters:} group-averaged most discriminant filters mapped to the sensor space, for $RMI$ and $LMI$ respectively. Although eight filters were used in the classification pipeline, for simplicity, only the filter corresponding to the most discriminant component for each condition is included. These values are normalized to compensate for differences between datasets, and signs are not considered, as they are irrelevant to our analysis. The resulting filters apply the highest weights to electrodes related to motor tasks on the corresponding contralateral side.
    For CSP and Riemannian methods, only the names of the ten nodes with the highest values are displayed.
    }
    \label{fig:state_methods}
\end{center}
\end{figure}

\begin{figure}[!htb]
\begin{center}
\includegraphics[width=1\linewidth]{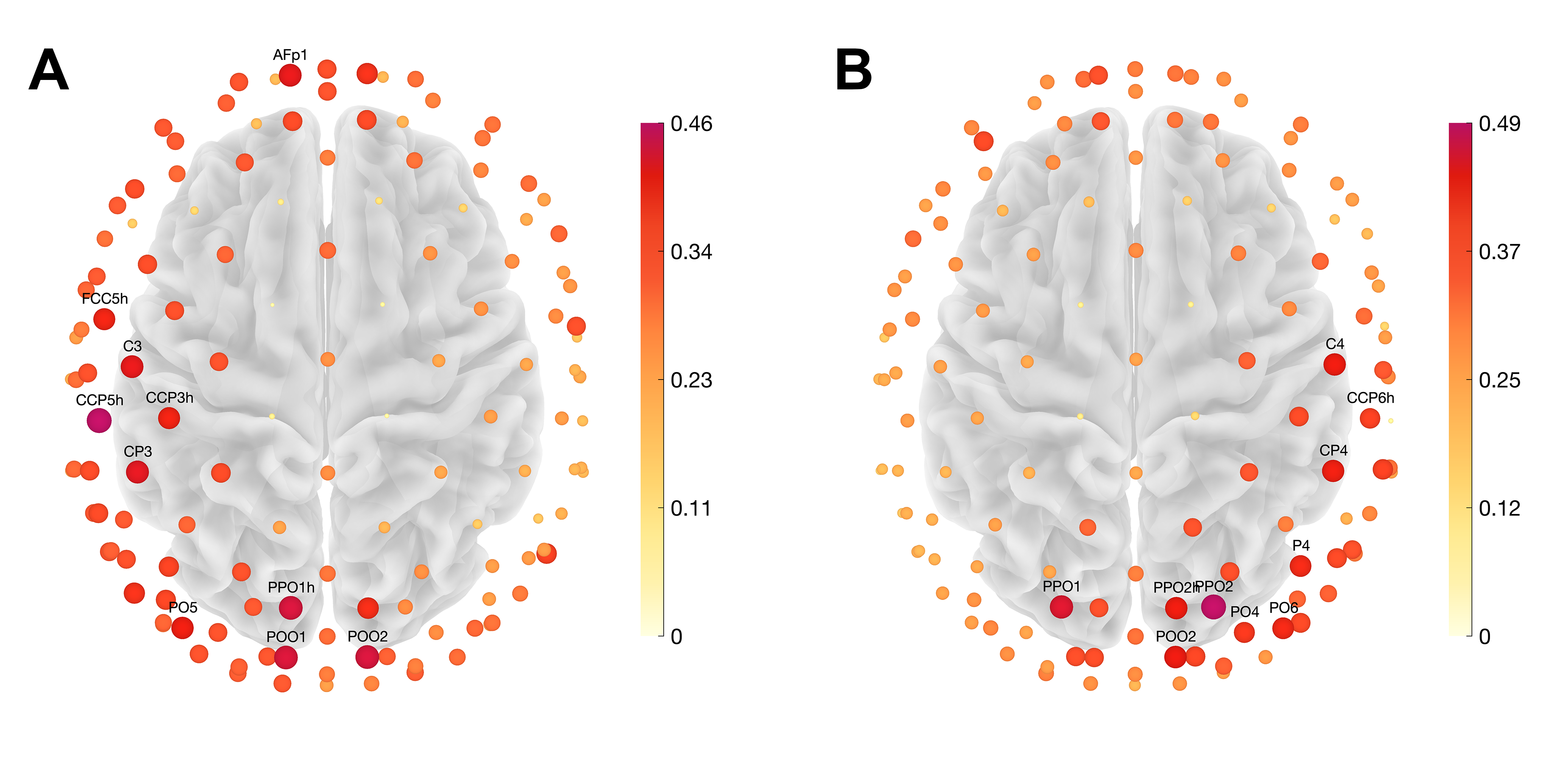}
    \caption[CSP patterns for $RMI$ and $LMI$.]{\textbf{CSP patterns for $RMI$ and $LMI$.}}
    \label{fig:csp_patterns}
\end{center}
\end{figure}

CSP filtering allows feature interpretation by examining the resulting filters and patterns. \textbf{Figure~\ref{fig:state_methods}C, D} and \textbf{Figure~\ref{fig:csp_patterns}} display the interpolation to sensor space of the group-averaged filters and patterns that best minimize each class's variance. For each subject, we included the absolute normalized topographic maps to provide a clear visualization of these features.

\begin{sidewaysfigure}[h!]
\begin{center}
\includegraphics[width=1\linewidth]{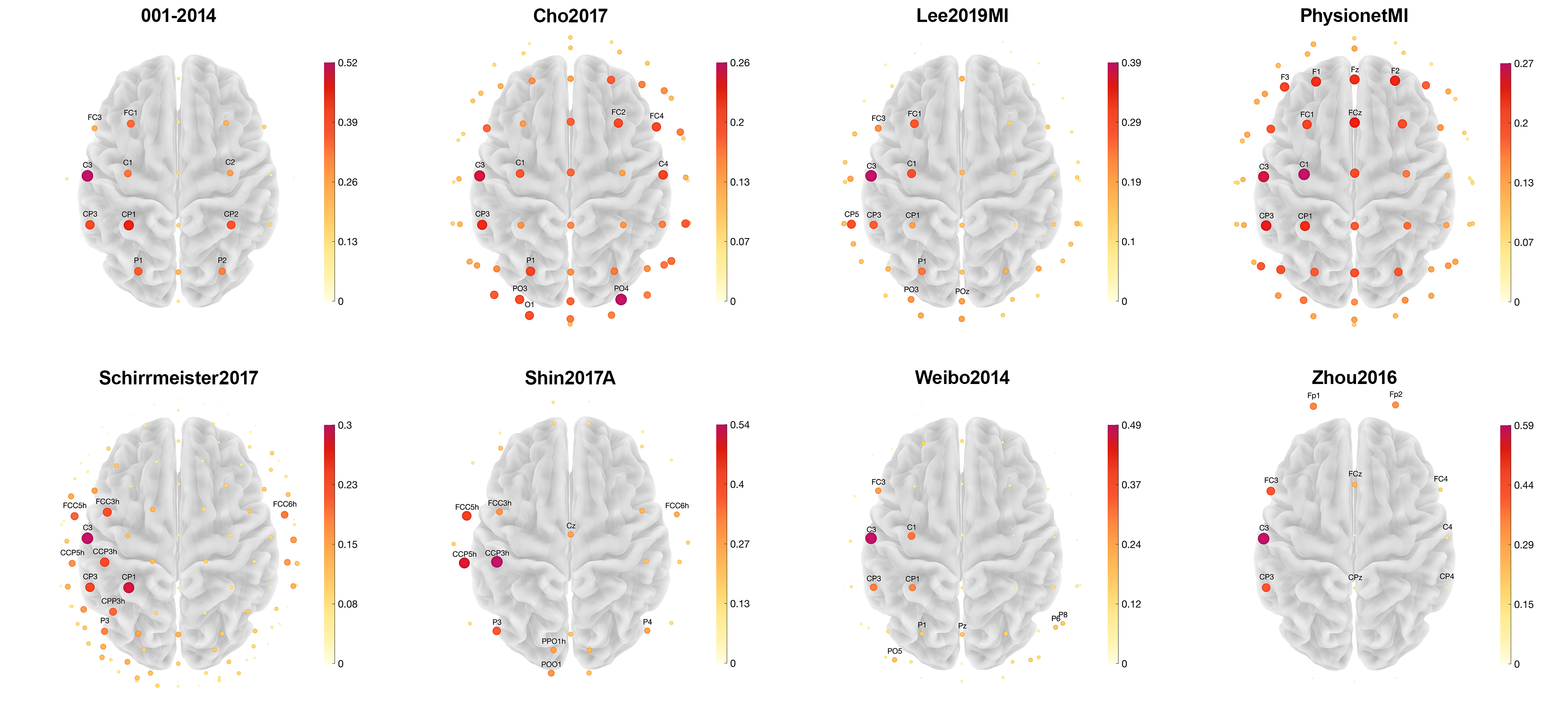}
    \caption[CSP.]{\textbf{CSP.} filters in $RMI$ per dataset.}
    \label{fig:sup_csp_filters_rh_all_dt}
\end{center}
\end{sidewaysfigure}

\begin{sidewaysfigure}[h!]
\begin{center}
\includegraphics[width=1\linewidth]{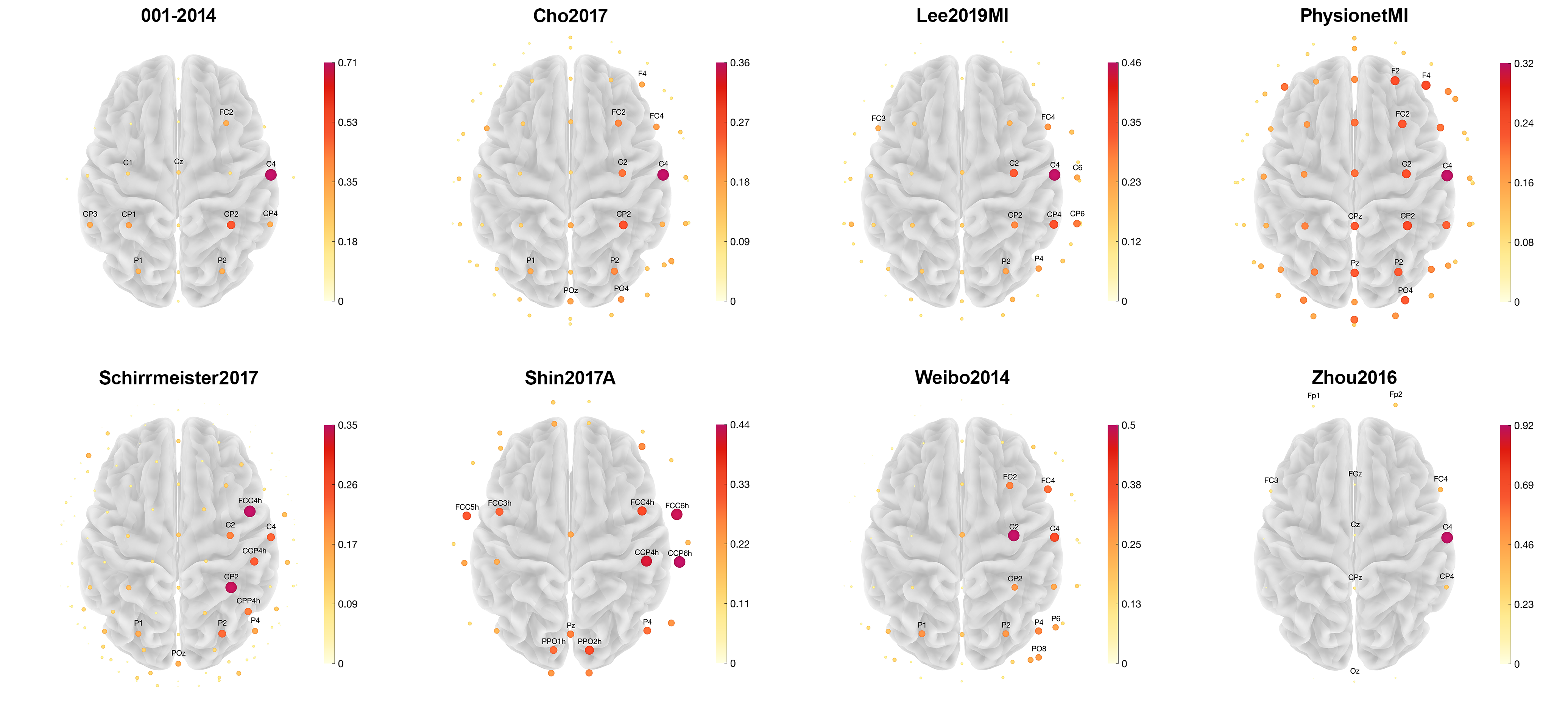}
    \caption[CSP.]{\textbf{CSP.} filters in $LMI$ per dataset.}
    \label{fig:sup_csp_filters_lh_all_dt}
\end{center}
\end{sidewaysfigure}

In $RMI$ (\textbf{Figure~\ref{fig:state_methods}C}), the maximum filter weights applied to electrodes corresponding to regions involved in hand-MI, specifically the contralateral M1, S1, and S2 cortices. C3 and its neighboring channels were the most highlighted by the $RMI$ filters across all datasets  (\textbf{Figure~\ref{fig:sup_csp_filters_rh_all_dt}}). The expected opposite behavior was observed for $LMI$ (\textbf{Figure~\ref{fig:state_methods}D}), with C4 and its surrounding electrodes obtaining the highest values (\textbf{Figure~\ref{fig:sup_csp_filters_lh_all_dt}}). This pattern is also consistent with the derived patterns (\textbf{Figure~\ref{fig:csp_patterns}}). Nonetheless, both filters and patterns showed involvement of parieto-occipital areas, which are primarily related to the associative visual cortex.

It is important to point out that CSP is not a source separation or localization method \cite{blankertz2007optimizing}. Instead, each filter is optimized to maximize the variance of one class while minimizing the variance of the other. In the context of $RMI$ versus $LMI$ paradigm, if we consider a filter that maximizes variance for $LMI$ class and minimizes it for $RMI$, then an expected high weight on the left hemispherical motor area can have two plausible causes. It can either originate from an ERD during $RMI$, or from an ERS during $LMI$ (where $RMI$ areas become more relaxed when focus is on $LMI$, leading to an increase in idle rhythm). It could also be a combination of both effects. Despite this potential ambiguity, the mixing effect is irrelevant for the discrimination task, although it presents a significant limitation for neurophysiological interpretation.

\subsubsection{Riemannian Space}
\begin{sidewaysfigure}[h!]
\begin{center}
\includegraphics[width=1\linewidth]{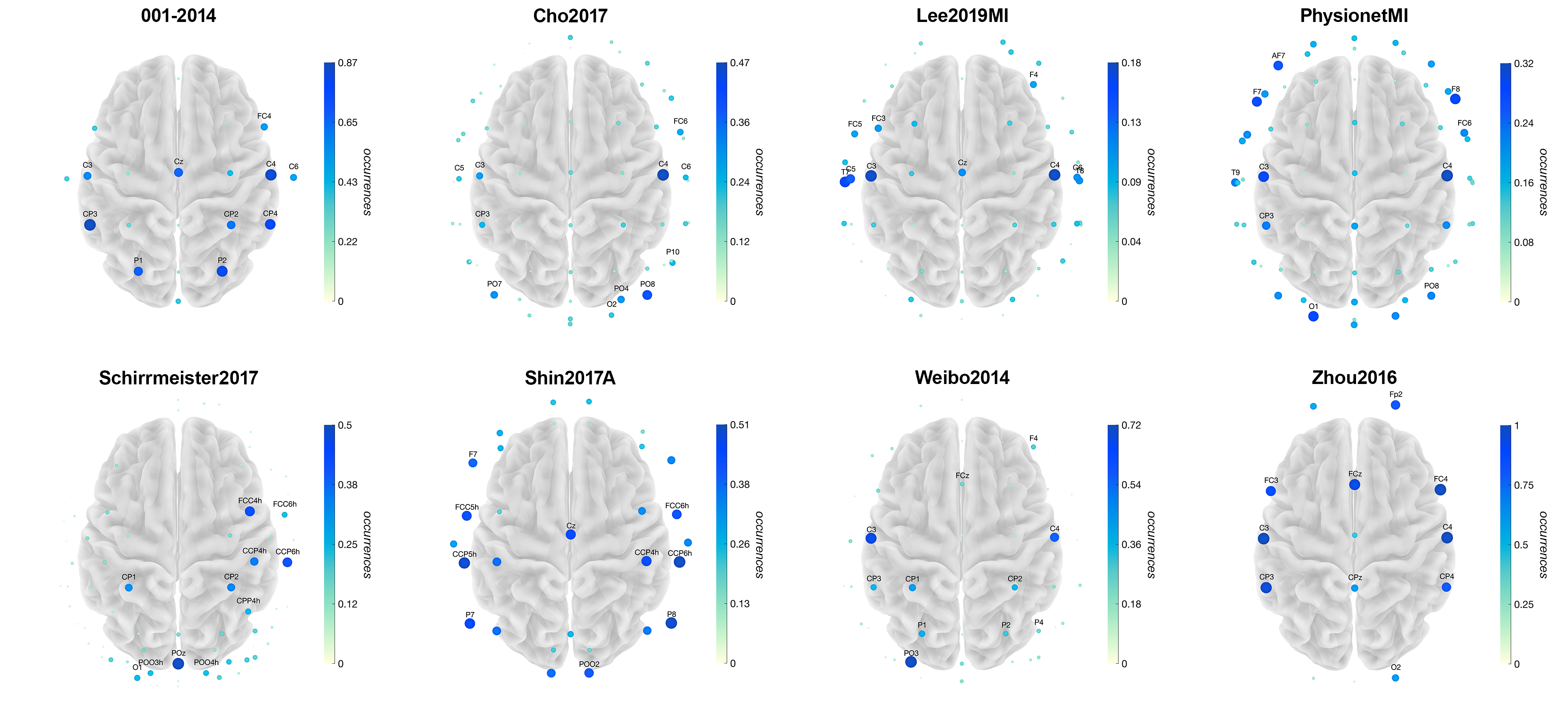}
    \caption[Riemannian.]{\textbf{Riemannian.} Feature occurrences per dataset.}
    \label{fig:sup_occurrences_riemannian_all_dt}
\end{center}
\end{sidewaysfigure}

We performed a channel selection in the manifold to validate the interpretation of Riemannian features. Within the 5-fold CV framework, the backward selection procedure identified the 10 channels that best maximized the Riemannian distance between classes for each subject. \textbf{Figure~\ref{fig:state_methods}B} summarizes the group-cumulative occurrences in a sensor plot. For each electrode, the number of selection times was normalized by the maximum possible occurrences. For instance, C4 was the most frequently selected electrode, with an occurrence rate of 50\% over the total possible selection times.
In general, we observed a concentration of features in the M1, S1, and S2 regions on both hemispheres, with a higher number of electrodes on the right hemisphere. These channels proved that Riemannian features were directly associated with the sensorimotor cortex. Nonetheless, we observed a subset of channels located in the parieto-occipital area that were not strictly related to MI. This behavior was consistently reproduced across almost all datasets (see \textbf{Figure~\ref{fig:sup_occurrences_riemannian_all_dt}}).

Even though this manipulation brings the Riemannian method closer to feature interpretation, it still lacks a clear understanding of which features are associated with each class. For instance, it is not possible to determine if occurrences in the motor cortex are related to the contralateral hand MI or if they result from the bilateral recruitment of these areas. Additionally, we observed a bias towards selections in the right hemisphere. One might speculate that this is due to higher resource consumption in the non-dominant hemisphere, correlating with results obtained from previous methods. However, this remains speculative and cannot be confirmed solely by examining the Riemannian selection.

\section{Discussion}
{\label{Discussion}}

This study was motivated by the hypothesis that brain network properties might have a beneficial role in distinguishing between different mental states relevant to BCIs. Specifically, we hypothesized that integrating the spatial component of hand-MI into the mathematical formulation of the network metrics might yield more interpretable and possibly more accurate results as compared to state-of-the-art-methods.

The results highlight that brain network lateralization serves as a distinctive feature for uncovering the underlying brain connectivity mechanisms in hand-MI, rendering it well-suited for classification purposes. We assessed the reproducibility of these findings across 288 subjects using an open-access toolkit \cite{jayaram2018moabb}, affirming its reliability. However, while our approach achieved competitive accuracies comparable to the traditional method, PSD, it did not attain the exceptional accuracies demonstrated by CSP and Riemannian methods.

\subsection{BCI performance and classification accuracy}
The promising discriminant network lateralization patterns were reflected in classification performance, yielding competitive scores compared to the benchmark established by PSD. This suggests that network-based features can capture essential aspects of brain activity that are relevant for distinguishing between mental states. However, neither of these methods achieved the scores obtained by CSP and Riemannian approaches. To understand the underlying reasons for this outcome, several important considerations should be noted.

First, it is important to recognize that our method does not include manipulations specifically designed to enhance classification performance, unlike CSP and Riemannian methods. For instance, CSP aims to directly maximize the variance ratio between two conditions rather than purely identifying the neural sources that generate that variance. Additionally, CSP is known to be sensitive to outliers; a single trial with high variance can significantly influence the resulting filters, potentially leading to unreliable outcomes \cite{blankertz2007optimizing}.

Secondly, while Riemannian methods are renowned for their accuracy, their primary limitation lies in their lack of interpretability. Many implementations based their results on Minimum Distance Mean (MDM), calculating distances between class mean SPD matrices without providing intermediate insights into the features driving these distances \cite{barachant2011multiclass}. Another widely used technique consists in projecting onto the tangent space. Here, careful attention must be paid to the dimensionality of SPD matrices, as features derived from high-dimensional covariance matrices are prone to overfitting due to the typically limited number of trials in BCI datasets \cite{lotte2007review, rodrigues2017dimensionality}. Although not the most widely adopted approach, the Riemannian method used in this study \cite{barachant2011channel} addresses both issues by selecting a restricted set of sensors based on their discriminative power, which facilitates subsequent interpretation of the results.

Recent publications have demonstrated their interest in validating Riemannian-based accuracies alongside comprehensive neurophysiological interpretations. \cite{larzabal2021riemannian} adopted a similar Riemannian selection approach as used in our project \cite{barachant2011channel}. But instead of working with accumulated occurrences, they attributed Riemannian distances to electrodes. Within each backward iteration, they assigned the Riemannian distance between classes to the electrode being removed, thereby establishing an inverse relationship between distance and the electrode's contribution to class separation. In line with our findings, they reported improved interpretability and classification performance compared to CSP. Additionally, other authors have explored combining CSP and Riemannian methods to enhance interpretative insights. For instance, \cite{xu2020tangent} studied spatial filters in the tangent space that enabled CSP-like pattern analysis while improving accuracies. Their approach resulted in patterns less susceptible to artifacts and capable of extracting additional neurophysiological activity compared to traditional CSP methods.

\subsection{Interpretability of BCI controlling features}
Our primary contribution lies in demonstrating that brain network lateralization properties can elucidate the underlying mechanisms of hand MI, thereby transforming them into promising features for MI-BCI systems. The relevance of these features was underscored by the spatial distribution of the most discriminant nodes, predominantly encompassing sensorimotor-related areas, thus ensuring that artifactual sources did not confound the differentiation.

The $t$-values topographical maps revealed that electrodes exhibiting the highest discriminant $strength$ were located in the contralateral hemisphere relative to the imagined movement. This observation highlights that lateralization manifests in the connectivity patterns of MI organization. Considering the anatomical symmetry of sensorimotor areas, this finding promoted the development of properties that compare functional lateralization across homotopic brain regions.

Interestingly, each introduced network metric emphasized different groups of nodes corresponding to distinct stages of the motor task. The $laterality$ $index$ specifically highlighted differences in nodes associated with motor planning and execution areas. Notably, this metric treats both hemispheres as isolated modules, excluding inter-hemispheric links. Consequently, it suggests that areas primarily dedicated to pure motor tasks rely predominantly on intra-hemispheric connections. This functional asymmetry supports the notion that homotopic brain regions exhibiting functional lateralization tend to have weaker inter-hemispheric connections, potentially enhancing the efficiency of processing lateralized functions \cite{karolis2019architecture}. Nonetheless, inter-hemispheric connections play a crucial complementary role in the complexities of motor tasks. For instance, coordinating reaching and grasping may require interactions between the contralateral and ipsilateral hemispheres \cite{gazzaniga2000cerebral}. Indeed, the inclusion of inter-hemispheric connections through $integration$ increased differentiation not only in areas related to motor execution but also in spatial sensory processing and information integration \cite{vanderah2020nolte, dehaene1998neuronal}. Furthermore, the contrast between intra- and inter-hemispheric connections through $segregation$ had the highest impact on areas typically involved in motor planning and other higher-order cognitive functions like decision-making and problem-solving \cite{martel2023tms, gerardin2000partially, gao2011evaluation, hetu2013neural}.

When compared to standard reference methods, PSD differentiation, much like integration, consistently highlighted electrodes over motor execution and sensory integration cortices. However, the expected inverse contralateral pattern was not observed, as significant PSD differences were only noted in the right hemisphere. This indicates a stronger response of PSD for $LMI$ or a bilateral activation \cite{sabate2004brain, zapala2020effects}.

For CSP, the spatial filters and patterns were well-aligned with the MI task, assigning the highest weights to electrodes over the contralateral sensorimotor cortex. However, the analysis also revealed significant influence from channels not typically associated with MI, particularly in the parieto-occipital region. Despite not being directly involved in MI, these areas might contribute due to their role in visual processing and spatial orientation \cite{wolbers2003contralateral, monaco2020decoding}, which can be inadvertently engaged during the MI task.

A similar trend emerged with the Riemannian-based method, exhibiting a concentration of relevant occurrences over the sensorimotor areas along with an unexpected subset of nodes in the parieto-occipital area. Intriguingly, this pattern was not detected by PSD or network analysis, methods focused solely on brain signal characterization rather than task differentiation. Indeed, Riemannian methods do not facilitate the association of specific patterns with each mental task, as their primary goal is to maximize the distance between SPD matrices rather than to characterize individual mental states. This discrepancy underscores the challenges in isolating pure MI features and highlights the necessity for refining techniques to mitigate contributions from unrelated channels.

\subsection{Methodological considerations}
Even though we demonstrated the reliability of our approach in identifying consistent neurophysiological sources across a considerable number of datasets, this study presents clear caveats that need to be acknowledged and addressed in the future. A first limitation is related to the signal preprocessing steps included in our pipeline. Indeed, only pass-band filtering was included. Incorporating additional filtering techniques, such as Common Average Reference (CAR) for re-referencing or Independent Component Analysis (ICA) for artifact suppression, could have been beneficial for this study and might have enhanced accuracy \cite{bashashati2007survey}. Nonetheless, by looking at the topographical $t$-test scalp maps on the network side, we can affirm that artifacts do not show a leading role in the results. 

With respect to pass-band filtering, we have worked with the assumption that MI generates distinguishing ERD/ERS in the $\alpha$ and $\beta$ frequency bands. But a more thorough study within sharper bands may be worthwhile since $\alpha$ and $\beta$ components differ with temporal behavior. \cite{pfurtscheller2001motor} have demonstrated the existence of at least three different types of oscillations at the same electrode location over the sensorimotor cortex in voluntary hand movement. Then working at different band levels may generate different and potentially more precise results. One possibility is to test the characteristics of each frequency band before immersing into feature extraction \cite{corsi2022functional} or work it out at the feature selection level by looking at precise single frequency bins \cite{cattai2021phase}. 

With respect to pass-band filtering, we have worked under the assumption that MI generates distinguishing ERD/ERS in the $\alpha$ and $\beta$ frequency bands. However, a more thorough investigation within narrower bands may be worthwhile since $\alpha$ and $\beta$ components differ in their temporal behavior. Indeed, it has already been demonstrated that there are at least three different types of oscillations at the same electrode location over the sensorimotor cortex during voluntary hand movement \cite{pfurtscheller2001motor}. Therefore, analyzing different band levels might yield different and potentially more precise results. One approach could be to test the characteristics of each frequency band before proceeding to feature extraction \cite{corsi2022functional}, or to refine the analysis at the feature selection stage by examining precise single frequency bins \cite{cattai2021phase}.

The lateralized nature of the hand-MI task studied here allowed us to effectively leverage this property by capturing it at the network level. However, it is important to acknowledge that FC alone already encodes the coordinated activity patterns underlying MI tasks \cite{brunner2006online, hamner2011phase, zhang2012improved, li2016decoding, feng2020functional}. Unlike network metrics, which simplify complex network characteristics into single values, FC provides a more nuanced perspective by examining pairwise interactions and depicting the intensity of connections between all pairs of electrodes, thereby offering rich data for feature extraction. 
Nevertheless, this advantage also poses challenges, notably the high-dimensionality of FC matrices, which can lead to potential overfitting. To address this issue, two viable strategies include channel selection and dimensionality reduction techniques. For instance, \cite{li2016decoding} applied principal component analysis (PCA) to concatenated FC matrices to reduce feature dimensionality. Their consistent high-performance results on the \textit{001-2014} dataset (82\%, only 2\%  below the winning accuracy reported by the competition \cite{blankertz2001classifying}) underscore the competitive potential of direct classification at the FC level. This highlights the promise of FC in BCI applications while emphasizing the need for careful consideration of data dimensionality.

Lastly, all these findings presuppose an approximate mapping between EEG channel locations and the corresponding underlying brain areas. Conducting further analysis in the source space could be advantageous for providing a more precise depiction of the neural mechanisms detected by our method \cite{jatoi2014survey, barzegaran2017functional}. However, addressing this approach involves considering two main limitations. Firstly, obtaining individual magnetic resonance images (MRIs) is crucial for creating a realistic brain model, but these data were unavailable for the datasets under study. Secondly, FC estimations can be sensitive to signal transformations, and outcomes may significantly vary based on the chosen reconstruction algorithm. Therefore, future research is essential to explore the robustness and consistency of our results at the source space level.

\section{Conclusion}

In this study, we assess the interpretability of the brain features used in MI-BCI tasks. In particular, we evaluated network-based metrics derived from brain FC. We hypothesized that the embedded spatial organization of the brain plays a pivotal role in discerning different BCI-related mental states. Our findings underscored that brain network lateralization is a unique attribute in hand-MI, presenting an ideal foundation for classification scenarios. The ensemble of introduced lateralization indexes demonstrated efficacy in identifying the key components that intervene at different stages of MI. Moreover, we provided a meta feature interpretation analysis, comparing our approach against three established methods —PSD, CSP, and Riemannian geometry— revealing that not all discriminant features were strictly tied to the MI task. This prompts critical questions about the interpretability of classification performance and the extent to which these scores genuinely reflect neural processes underlying MI.

In the BCI community, there are high expectations for the advancement of tools capable of decoding mental states. Two major conditions must be simultaneously reached,, i.e., high accuracy and neurological plausibility. While our method has validated the latter, further research is necessary to enhance the former.

\section*{Acknowledgments}
The research leading to these results has received funding from the European Research Council (ERC) under the European Union Horizon 2020 research and innovation program (Grant Agreement $864729$). 
The content is solely the responsibility of the authors and does not necessarily represent the official views of any of the funding agencies.


\par\null

\begin{small}
\end{small}


\newpage

\begin{table}[!thb]
\caption[Dataset attributes]{\textbf{Dataset attributes.} Overview of all included datasets with EEG recordings in a left versus right hand MI paradigm. \#: number, sub: subjects, ch: channels.}
\centering
\begin{small}
    \begin{tabular}{c|ccccccc}
     \toprule
     Dataset
        & \#sub 
        & \#ch
        & \#trial/class
        & epoch[s]
        & \#sessions  
        & ref.\\\hline
        
         001-2014          & 9     & 22    &144    &4      &2      & \fontsize{8}{8}\selectfont \cite{tangermann2012review} \\
         Cho2017           & 49    & 64    &100    &3      &1      & \fontsize{8}{8}\selectfont \cite{cho2017eeg} \\
         Grosse-Wentrup    & 10    & 128   &150    &7      &1      & \fontsize{8}{8}\selectfont \cite{grosse2009beamforming} \\
         Lee2019 MI        & 54    & 62    &100    &4      &2      & \fontsize{8}{8}\selectfont \cite{lee2019eeg} \\
         Physionet         & 109   & 64    &23     &7      &1      & \fontsize{8}{8}\selectfont \cite{goldberger2000physiobank} \\ 
         Schirrmeister2017 & 14    & 128   &120    &4      &1      & \fontsize{8}{8}\selectfont \cite{schirrmeister2017deep} \\
         Shin2017A         & 29    & 30    &30     &10     &3      & \fontsize{8}{8}\selectfont \cite{shin2016open} \\
         Weibo2014         & 10    & 60    &80     &4      &1      & \fontsize{8}{8}\selectfont \cite{yi2014evaluation} \\
         Zhou2016          & 4     & 14    &160    &5      &3      & \fontsize{8}{8}\selectfont \cite{zhou2016fully} \\\hline
        \end{tabular}
    \end{small}
    \label{tab:datasets}
\end{table}

\begin{table}[!t]
\caption[Classification performances: average accuracies.]{\textbf{Classification performances:} Average accuracies across methods for each dataset. RG: Riemannian geometry method, $s$: $strength$, $\lambda$: $laterality$ $index$, $\sigma$: $segregation$, $\omega$: $integration$.}
\centering
\begin{small}
    \begin{tabular}{c@{\hskip 0.06in}|c@{\hskip 0.06in}c@{\hskip 0.06in}c@{\hskip 0.06in}c@{\hskip 0.06in}c@{\hskip 0.06in}c@{\hskip 0.06in}c@{\hskip 0.06in}}
     \toprule
        \fontsize{8}{8}\selectfont Dataset
        & \fontsize{8}{8}\selectfont PSD+SVM 
        & \fontsize{8}{8}\selectfont $s$+SVM 
        & \fontsize{8}{8}\selectfont $\lambda$+SVM
        & \fontsize{8}{8}\selectfont $\sigma$+SVM
        & \fontsize{8}{8}\selectfont $\omega$+SVM
        & \fontsize{8}{8}\selectfont CSP+SVM
        & \fontsize{8}{8}\selectfont RG+SVM\\\hline
        
        \fontsize{7}{7}\selectfont 001-2014            
                            &\fontsize{6.5}{6.5}\selectfont 73.5$\pm$11.4    
                            &\fontsize{6.5}{6.5}\selectfont 77.2$\pm$15.1 
                            &\fontsize{6.5}{6.5}\selectfont 77.0$\pm$14.5
                            &\fontsize{6.5}{6.5}\selectfont 73.2$\pm$16.3
                            &\fontsize{6.5}{6.5}\selectfont 77.6$\pm$14.1
                            &\fontsize{6.5}{6.5}\selectfont 85.6$\pm$12.8
                            &\fontsize{6.5}{6.5}\selectfont 85.3$\pm$13.0 \\
        
        \fontsize{7}{7}\selectfont Cho2017	
                            &\fontsize{6.5}{6.5}\selectfont 64.7$\pm$12.0	
                            &\fontsize{6.5}{6.5}\selectfont 64.5$\pm$12.0
                            &\fontsize{6.5}{6.5}\selectfont 64.1$\pm$11.6
                            &\fontsize{6.5}{6.5}\selectfont 61.2$\pm$10.6	
                            &\fontsize{6.5}{6.5}\selectfont 64.0$\pm$12.1
                            &\fontsize{6.5}{6.5}\selectfont 72.9$\pm$13.5 
                            &\fontsize{6.5}{6.5}\selectfont 75.2$\pm$11.9\\

        \fontsize{7}{7}\selectfont Grosse-Wentrup
                            &\fontsize{6.5}{6.5}\selectfont 75.3$\pm$14.3	
                            &\fontsize{6.5}{6.5}\selectfont 69.2$\pm$12.6	
                            &\fontsize{6.5}{6.5}\selectfont 75.0$\pm$13.4
                            &\fontsize{6.5}{6.5}\selectfont 71.6$\pm$12.5
                            &\fontsize{6.5}{6.5}\selectfont 70.5$\pm$13.1 
                            &\fontsize{6.5}{6.5}\selectfont 77.3$\pm$22.3 
                            &\fontsize{6.5}{6.5}\selectfont 80.9$\pm$19.5\\
                            
        \fontsize{7}{7}\selectfont Lee2019 MI	   
                            &\fontsize{6.5}{6.5}\selectfont 67.2$\pm$15.4
                            &\fontsize{6.5}{6.5}\selectfont 62.2$\pm$13.4	
                            &\fontsize{6.5}{6.5}\selectfont 63.3$\pm$14.9
                            &\fontsize{6.5}{6.5}\selectfont 60.8$\pm$11.7
                            &\fontsize{6.5}{6.5}\selectfont 64.8$\pm$15.0
                            &\fontsize{6.5}{6.5}\selectfont 74.7$\pm$18.3
                            &\fontsize{6.5}{6.5}\selectfont 76.1$\pm$17.9\\
                                                                                 
        \fontsize{7}{7}\selectfont Physionet	
                            &\fontsize{6.5}{6.5}\selectfont 61.4$\pm$15.2
                            &\fontsize{6.5}{6.5}\selectfont 60.2$\pm$13.2
                            &\fontsize{6.5}{6.5}\selectfont 63.3$\pm$15.1
                            &\fontsize{6.5}{6.5}\selectfont 62.5$\pm$14.0
                            &\fontsize{6.5}{6.5}\selectfont 62.4$\pm$16.2
                            &\fontsize{6.5}{6.5}\selectfont 70.1$\pm$17.4
                            &\fontsize{6.5}{6.5}\selectfont 73.7$\pm$16.1\\
                            
        \fontsize{7}{7}\selectfont Schirrmeister2017	
                            &\fontsize{6.5}{6.5}\selectfont 74.6$\pm$13.0
                            &\fontsize{6.5}{6.5}\selectfont 70.7$\pm$10.9
                            &\fontsize{6.5}{6.5}\selectfont 73.6$\pm$11.6
                            &\fontsize{6.5}{6.5}\selectfont 68.6$\pm$11.2
                            &\fontsize{6.5}{6.5}\selectfont 70.5$\pm$11.5	
                            &\fontsize{6.5}{6.5}\selectfont 83.1$\pm$14.9 
                            &\fontsize{6.5}{6.5}\selectfont 88.2$\pm$11.6\\

        \fontsize{7}{7}\selectfont Shin2017A	     
                            &\fontsize{6.5}{6.5}\selectfont 56.6$\pm$20.8
                            &\fontsize{6.5}{6.5}\selectfont 60.7$\pm$16.9
                            &\fontsize{6.5}{6.5}\selectfont 66.1$\pm$21.7
                            &\fontsize{6.5}{6.5}\selectfont 66.6$\pm$19.8
                            &\fontsize{6.5}{6.5}\selectfont 66.5$\pm$20.1
                            &\fontsize{6.5}{6.5}\selectfont 75.6$\pm$20.7
                            &\fontsize{6.5}{6.5}\selectfont 76.0$\pm$20.2\\
        
        \fontsize{7}{7}\selectfont Weibo2014	        
                            &\fontsize{6.5}{6.5}\selectfont 72.1$\pm$14.2	
                            &\fontsize{6.5}{6.5}\selectfont 71.5$\pm$14.3
                            &\fontsize{6.5}{6.5}\selectfont 70.0$\pm$14.4
                            &\fontsize{6.5}{6.5}\selectfont 62.3$\pm$12.5
                            &\fontsize{6.5}{6.5}\selectfont 72.7$\pm$15.6
                            &\fontsize{6.5}{6.5}\selectfont 83.1$\pm$14.4	
                            &\fontsize{6.5}{6.5}\selectfont 84.8$\pm$14.0\\
        
        \fontsize{7}{7}\selectfont Zhou2016	        
                            &\fontsize{6.5}{6.5}\selectfont 83.5$\pm$6.9
                            &\fontsize{6.5}{6.5}\selectfont 88.2$\pm$5.8
                            &\fontsize{6.5}{6.5}\selectfont 85.9$\pm$6.8
                            &\fontsize{6.5}{6.5}\selectfont 84.3$\pm$9.7
                            &\fontsize{6.5}{6.5}\selectfont 89.1$\pm$6.0
                            &\fontsize{6.5}{6.5}\selectfont 94.5$\pm$5.8 
                            &\fontsize{6.5}{6.5}\selectfont 94.4$\pm$5.6\\\hline

        \fontsize{7}{7}\selectfont MEAN	       
                            &\fontsize{6.5}{6.5}\selectfont 69.9$\pm$8.2
                            &\fontsize{6.5}{6.5}\selectfont 69.4$\pm$9.1
                            &\fontsize{6.5}{6.5}\selectfont 70.9$\pm$7.7
                            &\fontsize{6.5}{6.5}\selectfont 67.9$\pm$7.7
                            &\fontsize{6.5}{6.5}\selectfont 70.9$\pm$8.3
                            &\fontsize{6.5}{6.5}\selectfont 79.7$\pm$7.6 
                            &\fontsize{6.5}{6.5}\selectfont 81.6$\pm$7.0\\
        \hline
        \end{tabular}
    \end{small}
    \label{tab:classif_acc}
\end{table}

\begin{Boxes}[float=ht!, label=box1, title=Box 1 - Within and inter-hemisphere connections]
\noindent lateralization properties are based on FC within or across hemispheres. For a node $i$ in the left hemisphere, the within-hemisphere strength ($LL_{i}$) is measured by summing the connectivity values between it and the other nodes located in the left hemisphere. For EEG-based networks, this also includes the connections that node $i$ establishes with the central line electrodes ($LC_{i}$). In contrast, the across-hemisphere strength ($LR_{i}$) is estimated by summing the connectivity between node $i$ and all nodes located in the right hemisphere.

\begin{align}
  & LL_{i} =\sum_{l{\ne}i}^{L}W_{il(LL)},&
  & LC_{i} =\sum_{c}^{C}W_{ic(LC)},& 
  & LR_{i} =\sum_{r}^{R}W_{ir(LR)}
\end{align}

Similarly, for a node $j$ in the right hemisphere, $RR_{j}$, $RC_{j}$ and $RL_{j}$ are obtained using the same reasoning. 
\begin{align}
  & RR_{j} =\sum_{r{\ne}j}^{R}W_{jr(RR)},&
  & RC_{j} =\sum_{c}^{C}W_{jc(RC)},&  
  & RL_{j} =\sum_{l}^{L}W_{jl(RL)}
\end{align}

The same approach is applied to a node $k$ located in the EEG central line to obtain $CC_{k}$, $CR_{k}$ and $CL_{k}$. 
\begin{align}
  & CC_{k} =\sum_{c{\ne}k}^{C}W_{kc(CC)},&
  & CR_{k} =\sum_{r}^{R}W_{kr(CR)},&  
  & CL_{k} =\sum_{l}^{L}W_{kl(CL)}
\end{align}

For obvious reasons, the concepts of $segregation$ and $integration$ do not apply to nodes located on the central line.

\begin{center} 
\tcbox[left=0mm,right=-0mm,top=0mm,bottom=0mm,boxsep=0mm,arc=0mm,boxrule=0pt]{%
\includegraphics[width=12cm]{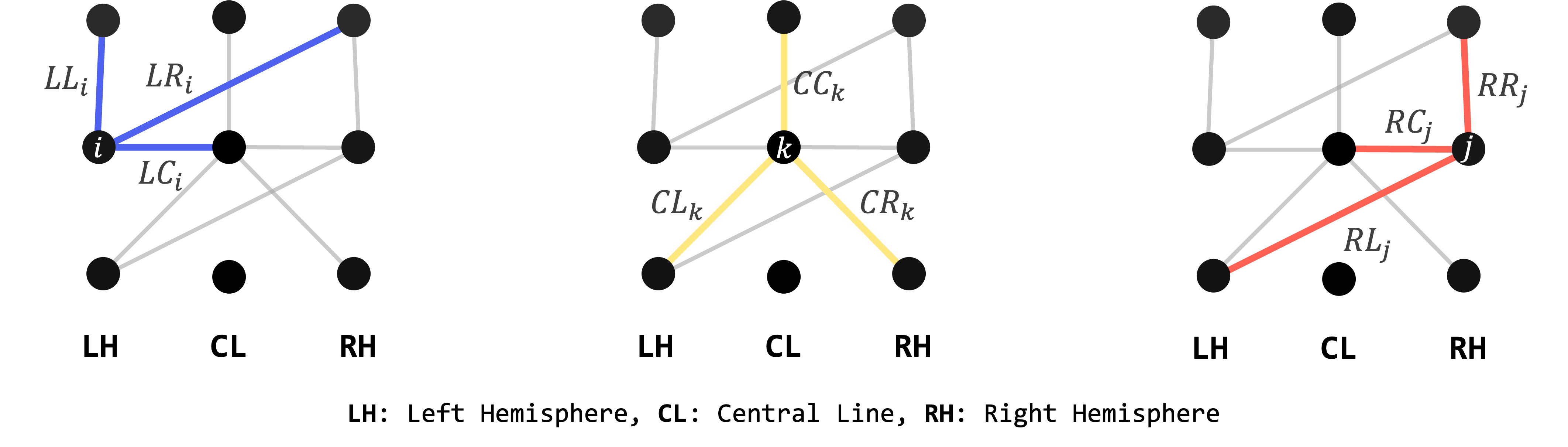}}
\end{center}

To clarify the notation, each capital letter term respectively denotes the locations of node $i$ and the nodes it connects with (e.g. $LR_{i}$ indicates that node $i$ belongs to the left hemisphere and considers the connections linking it to nodes in the right hemisphere).
\label{box:box1}
\end{Boxes}

\clearpage
\appendix
\beginsupplement
\section*{Appendix}
{\label{Appendix}}

\begin{figure}[h!]
\begin{center}
\includegraphics[width=1\linewidth]{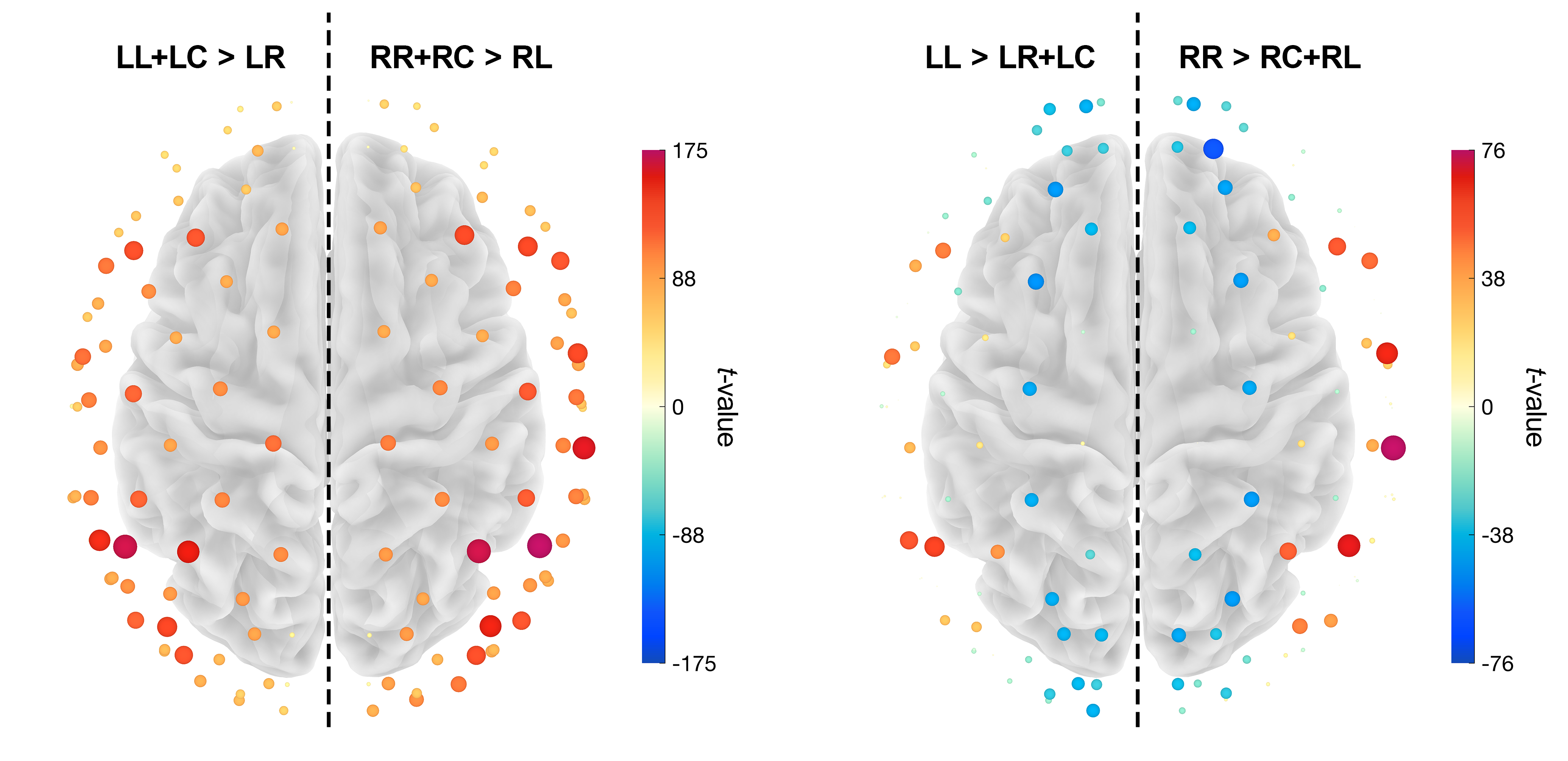}
    \caption[Influence of middle line links.]{\textbf{Influence of middle line links.} 
    Group average $t$-values that experimentally demonstrate the influence of middle line edges on each hemisphere. At each node, we statistically compare the within connection versus the inter-hemispheric in two possible scenarios. On the left, we show the results when we consider the influence of including middle line links ($LC_{i}$, $RC_{j}$) as within-hemisphere. On the right, the results of considering them as inter-hemispheric. Excluding $LC_{i}$ and $RC_{j}$ from the within-connections has a localized negative impact on nodes closer to the central line. Conversely, $LL_{i} + LC_{i} > LR_{i}$ and $RR_{j} + RC_{j} > RL_{j}$ ensure positives values for each hemisphere's segregation. Therefore, when analysing the lateralization of $\sigma$, a negative value indicates stronger segregation on the right hemisphere.
}
    \label{fig:t-test_LC_RC}
\end{center}
\end{figure}

\end{document}